\documentclass[11pt]{article}
\usepackage{amssymb,cite}
\usepackage{graphicx}
\usepackage{latexsym}
\usepackage{wasysym}
\def\thefootnote{\fnsymbol{footnote}}
\newcommand{\eq}{\begin{equation}}
\newcommand{\en}{\end{equation}}
\newcommand{\be}{\begin{equation}}
\newcommand{\ee}{\end{equation}}
\newcommand{\eqa}{\begin{eqnarray}}
\newcommand{\ena}{\end{eqnarray}}
\newcommand{\ba}{\begin{eqnarray}}
\newcommand{\ea}{\end{eqnarray}}

\newcommand{\bra}{\langle}

\newcommand{\ket}{\rangle}

\newcommand{\ZZ}{\hbox{{\rm Z{\hbox to 3pt{\hss\rm Z}}}}}

\newcommand{\Z}{\mathbb{Z}}
\newcommand{\Zp}{Z_{\mbox{\tiny{p}}}}
\newcommand{\Za}{Z_{\mbox{\tiny{a}}}}
\newcommand{\Ep}{E_{\mbox{\tiny{p}}}}
\newcommand{\Ea}{E_{\mbox{\tiny{a}}}}
\newcommand{\Es}{E_{\mbox{\tiny{s}}}}
\newcommand{\Tc}{T_{\mbox{\tiny{c}}}}

\newcommand{\cale}{\mathcal{E}}

\begin{document}
\begin{titlepage}
\vskip0.5cm
\begin{flushright}
DFTT 10/07\\
\end{flushright}
\vskip0.5cm
\begin{center}
{\Large\bf The interface free energy: Comparison of accurate Monte Carlo 
results for the 3D Ising model with effective interface models
 } 
\end{center}
\vskip1.3cm
\centerline{
Michele~Caselle$^{a}$, Martin~Hasenbusch$^{b}$
 and Marco~Panero$^{c}$}
 \vskip1.0cm
 \centerline{\sl  $^a$ Dipartimento di Fisica
 Teorica dell'Universit\`a di Torino and I.N.F.N.,}
 \centerline{\sl Via Pietro~Giuria 1, I-10125 Torino, Italy}
 \centerline{{\sl
e--mail:} \hskip 6mm
 \texttt{caselle@to.infn.it}}
 \vskip0.4 cm
 \centerline{\sl  $^b$  
     Dipartimento di Fisica
     dell'Universit\`a di Pisa and I.N.F.N.,} 
   \centerline{ \sl Largo Bruno~Pontecorvo 3, I-56127 Pisa, Italy}

 \centerline{{\sl
e--mail:} \hskip 6mm
 \texttt{Martin.Hasenbusch@df.unipi.it}}
\vskip0.4 cm
 \centerline{\sl  $^c$ Institute for Theoretical Physics,
University of Regensburg,}
 \centerline{\sl
                  93040 -- Regensburg,
                              Germany}
 \centerline{{\sl
e--mail:} \hskip 6mm
 \texttt{marco.panero@physik.uni-regensburg.de}}
 \vskip1.0cm
\begin{abstract}
We provide accurate Monte Carlo results for the free energy of interfaces with periodic boundary conditions in the 3D Ising model. We study a large range of inverse temperatures, allowing to control corrections to scaling. In addition to square interfaces, we study rectangular interfaces for a large range of aspect ratios $u=L_1/L_2$. Our numerical results are compared with predictions of effective interface models. This comparison  verifies clearly  the effective Nambu-Goto model up to two-loop order. Our data also allow us to obtain the estimates  $\Tc/\sqrt{\sigma}=1.235(2)$, $m_{0++}/\sqrt{\sigma}=3.037(16)$ {\sl and $R_+=f_+^2 \sigma_0=0.387(2)$}, which are more precise than previous ones.
\end{abstract}
\end{titlepage}

\setcounter{footnote}{0}
\def\thefootnote{\arabic{footnote}}

\section{Introduction}
\label{introsect}

Interfaces play an important r\^ole in various fields of natural sciences. In soft condensed matter physics, in chemistry and in biology, interfaces separating two different media, for instance two different magnetization domains, or two different fluids, or a fluid and its vapour, are studied. The properties of such interfaces might be described by a unique effective model such as the capillary wave model~\cite{Privman:1992zv}.

Our motivation to study interfaces originates from the theory of high energy physics. An interface with given boundary conditions can be associated with the world-sheet of a fluctuating flux tube in the confinement regime of a gauge theory.
For intermediate and long distances between the sources, the relevant degrees of freedom for a system of confined quarks are supposed to be independent of the short distance gauge interaction, and might be modelled by string fluctuations (\emph{effective string picture}).

The simplest set-up for a numerical study of interfaces is provided by the Ising spin model on a simple cubic lattice. Its duality with respect to the $\Z_2$ gauge model~\cite{Wegner} maps the ordered phase to the confined regime. 

The classical Hamiltonian of the Ising spin model reads:
\begin{equation}
H(\{J\},\{h\},\{s\}) = - \sum_{ \bra xy \ket } J_{\bra xy \ket}  s_x s_y \; 
- \sum_x h_x s_x \; \;, \;\;\; s_x \in \{ 1, -1 \} \;\; ,
\end{equation}
where $x=(x_0,x_1,x_2)$ is a site of the lattice, and $\bra xy \ket$ denotes a pair of nearest neighbours on the lattice.  Here and in the following, the lattice spacing $a$ is set to $1$, and we shall always consider the case of a vanishing external field $h_x=0$, $\forall x$. The site coordinates run over $ 0 \le x_i \le L_i - 1$, where $i \in \{0,1,2\}$ label the three directions. 

In the case of periodic boundary conditions we take $J_{\bra xy \ket}=1$ for all links $ \bra xy \ket$. Anti-periodic boundary conditions, say, in the direction $0$, can be implemented imposing $J_{\bra xy \ket}=-1$ if $x=(L_0-1,x_1,x_2)$ and $y=(0,x_1,x_2)$, and $J_{\bra xy \ket}=1$ otherwise. 

The partition function is obtained as the sum over all configurations $\{s\}$ of the Boltzmann factor: 
\begin{equation}
 Z_{\{J\}}(\beta) = \sum_{\{s\}} \exp\left(-\beta H(\{J\},\{s\})\right) \;\;,
\end{equation}
where $\beta=1/(k_{\mbox{\tiny{B}}} T)$ is the inverse of the temperature of the three-dimensional classical spin model.

The goal of our work is to study an interface between the phases of positive and negative magnetization in the low-temperature regime of the spin model --- which corresponds to the confining regime of the gauge theory. 

Such an interface can be forced into the system by appropriate boundary conditions. For instance, one could constrain the spins at  $x_0=0$ to take the value $-1$ and those at $x_0=L_0-1$ to $+1$; here, however, we use anti-periodic boundary conditions in $0$-direction, because the finite $L_0$ effects are smaller and better understood than for Dirichlet boundary conditions.

In recent works~\cite{Caselle:2005vq, Panero:2005iu, Caselle:2005xy, Panero:2004zq, Caselle:2004jq, Caselle:2003db, Caselle:2003rq, Caselle:2002ah, Caselle:2002vq, Caselle:2002rm} we studied interfaces with Dirichlet boundary conditions in one direction and periodic boundary conditions in the other direction: via duality, this corresponds to a Polyakov loop correlator in the gauge model.

The comparison with the Nambu-Goto effective string model resulted in unexpected discrepancies at subleading orders. While finite $L_2$ corrections, in the direction with periodic boundary conditions, are described well by the effective theory, the finite $L_1$ corrections, in the direction of the Dirichlet boundary conditions, show unexpected deviations.

In order to further investigate this issue, we pick up again the work on interfaces with periodic boundary conditions in both directions. In~\cite{Caselle:2006dv, Billo:2005ej, Billo:2006zg} such a comparison had been performed for square interfaces $L_1=L_2$; in these studies, the numerical values of the interface tension were taken from~\cite{Caselle:2004jq}. 

In the present work, our results for the interface free energy allow for an independent determination of the interface tension, which is computed in technically quite a different way with respect to~\cite{Caselle:2004jq}; the consistency of the two results provides a non-trivial check of their validity. 

We obtain results for a large range of the inverse temperature $\beta$, allowing to study possible scaling corrections. Furthermore, we also compute the interface free energy for $L_1 \ne L_2$ for a large range of $u=L_1/L_2$: this enables us to compare with the non-trivial dependence on $u$, which is predicted by the effective interface models.

Finally, the results for the interface tension are also used in combination with a series analysis of the second moment correlation length in the high temperature phase. This yields a precise estimate of the universal amplitude ratio: 
\begin{equation}
 R_{+} = f_{\mbox{\tiny{2nd,+}}}^2 \sigma_0 \;\; ,
\end{equation}
where the amplitudes are defined by $\sigma \simeq  \sigma_0 (-t)^{\mu}$ and $\xi_{\mbox{\tiny{2nd}}} \simeq f_{\mbox{\tiny{2nd,+}}} t^{-\nu}$. Here, $\sigma$ is the interface tension, $\xi_{\mbox{\tiny{2nd}}}$ the second moment correlation length in the high temperature phase,  $t=(T-\Tc)/\Tc$ the reduced temperature and $\nu$, $\mu=2\nu$ the critical exponents of the correlation length and the interface tension, respectively. The result for $R_{+}$ can be compared e.g. with results obtained from experiments on binary mixtures.

We also update the estimate for:
\begin{equation}
 m_{0++} /\sqrt{\sigma} \; ,
\end{equation}
where now the error is dominated by the estimate of the mass $m_{0++}$ of the $0^{++}$ glueball. Note that under duality the interface tension of the Ising spin model is equal to the string tension of the  $\Z_2$ gauge model and the exponential correlation length in the low temperature phase of the Ising spin model is equal to the inverse mass of the $0^{++}$ glueball in confined phase of the $\Z_2$ gauge model. While there is no direct experimental particle  physics relevance of this result, it is interesting for theoretical reasons to compare $m_{0++} /\sqrt{\sigma}$ obtained from different gauge theories. Finally we also provide an updated estimate of the finite temperature transition $\Tc/\sqrt{\sigma}$. Note that here we refer to the temperature of the two-dimensional quantum  field theory. Its temperature is given by $T=1/(a L_0)$ and should not be confused with the temperature of the three-dimensional classical system defined above. In the following we shall denote the critical value of $L_0$ by $N_t$; i.e. $\Tc=1/(a N_t)$.

The content of this paper is the following: In section~\ref{freeenergydefinitionsect} we define  the interface free energy for finite interface area $L_1 \times L_2$ and finite transverse size of the system $L_0$. Next, in section~\ref{effectivestringsect}, we briefly summarize the predictions for the dependence of the interface free energy on $(L_1,L_2)$, according to an effective string-like description. In section~\ref{methodsect} 
we present our numerical method to compute the interface free energy. Our results for square and for rectangular interfaces are presented in section~\ref{interfacesect}, while section~\ref{amplituderatiosect} contains our results for the universal amplitude ratios. A summary and our conclusions are given in section~\ref{conclusionsect}. The numerical integration methods are presented in the appendix~\ref{integrationschemesect}.

\section{Definition of the interface free energy}
\label{freeenergydefinitionsect}

The basic quantity that we shall determine numerically is the ratio between the partition functions of the system with anti-periodic $\Za$ and periodic boundary conditions $\Zp$. The purpose of this section is to provide a definition of the
interface free energy in terms of this ratio.

The ratio $\Za/\Zp$ can be expressed in terms of the eigenvalues 
$\lambda_{n\mbox{\tiny{x}}}$ of the transfer matrix and the 
parity $p_{n\mbox{\tiny{x}}}= \pm 1$ of its eigenstates:
\footnote{Eq.~(\ref{evandpar}) can be justified as it follows: 
In the basis of slice configurations $\Sigma$, the matrix associated 
with anti-periodic boundary conditions is given by 
$P_{\Sigma',\Sigma} = \delta_{\Sigma',-\Sigma}$, where $-\Sigma$ means that 
all spins in the slice are flipped. Since the external field $h$ is vanishing,
the transfer matrix commutes with $P_{\Sigma',\Sigma}$; 
furthermore, $P$ squares to the identity, therefore it has eigenvalues 
$p_{n\mbox{\tiny{x}}} = \pm 1$. We label eigenvectors with $p_{n\mbox{\tiny{x}}} = 1$ by x$=$s and those with $p_{n\mbox{\tiny{x}}} = -1$ by x$=$a. Following standard conventions $\lambda_{n \mbox{\tiny{x}}}$ is decreasing with increasing 
$n$. 
}
\begin{equation}
\label{evandpar}
\frac{\Za}{\Zp} = \frac{\sum_{n} \sum_{x=s,a} p_{n \mbox{\tiny{x}}} \lambda_{n \mbox{\tiny{x}}}^{L_0}}
                       {\sum_{n} \sum_{x=s,a} \lambda_{n \mbox{\tiny{x}}}^{L_0}} \;\;.
\end{equation}
For $L_0 \gg \xi$, where $\xi=- 1/\ln(\lambda_{1\mbox{\tiny{s}}}/\lambda_{0\mbox{\tiny{s}}})$ is the bulk correlation length or the inverse of the mass of the 
theory, 
the partition function ratio in eq.~(\ref{evandpar}) is dominated by the largest eigenvalues $\lambda_{0\mbox{\tiny{s}}}$ and $\lambda_{0\mbox{\tiny{a}}}$:
\begin{equation}
 \frac{\Za}{\Zp} \simeq 
\frac{\lambda_{0\mbox{\tiny{s}}}^{L_0}-\lambda_{0\mbox{\tiny{a}}}^{L_0}}
{\lambda_{0\mbox{\tiny{s}}}^{L_0}+\lambda_{0\mbox{\tiny{a}}}^{L_0}}
= 
\frac{1-(\lambda_{0\mbox{\tiny{a}}}/\lambda_{0\mbox{\tiny{s}}})^{L_0}}{1+(\lambda_{0\mbox{\tiny{a}}}/\lambda_{0\mbox{\tiny{s}}})^{L_0}} \; .
\end{equation}
In this regime, the so-called tunneling mass: 
\begin{equation}
\label{tunnelmass}
m_{\mbox{\tiny{t}}} = - \ln(\lambda_{0\mbox{\tiny{a}}}/\lambda_{0\mbox{\tiny{s}}})
\end{equation}
can thus be obtained from: 
\begin{equation}
\label{massandratio}
m_{\mbox{\tiny{t}}} = - \frac{1}{L_0} \ln \left(\frac{1-\Za/\Zp}{1+\Za/\Zp} \right) \; .
\end{equation}
Now let us relate the ratio of partition functions with the phenomenological picture of interfaces separating the phases of positive and negative magnetisation. We assume that, to the leading approximation, the free energy of an interface is proportional to its area. Hence, for finite $L_0$, in the $L_1,L_2 \rightarrow \infty$ limit, 
there is only one interface in the system with anti-periodic boundary conditions and none in the system with periodic boundary conditions. Based on this scenario, the interface free energy is  naturally defined as: 
\begin{equation}
\label{fs1}
 F_{\mbox{\tiny{s}}}^{(1)} = -\ln(\Za/\Zp) + \ln L_0 \;\; ,
\end{equation}
where the $\ln L_0$ term takes into account the ``entropy'' due to the fact that the interface can be located at any point in the $x_0$-direction.\footnote{In principle, one might also add a further $\ln 2$ term, to take into account that the positive magnetization domain can be realized on the left-hand side of the interface and the negative one on its right-hand side, or \emph{vice versa}.}

Note that for finite $L_1,L_2$ the value of $F_{\mbox{\tiny{s}}}^{(1)}$ 
depends on $L_0$ and in particular, the limit $L_0 \rightarrow \infty$ 
is not finite. This last problem is related to the fact that for sufficiently 
large $L_0$, it is favoured by the entropy to create additional pairs of 
interfaces.

The presence of additional pairs of interfaces can be addressed in the dilute
gas approximation. I.e. we assume that the interaction of two interfaces
is short ranged and that the average distance between interfaces is large 
compared with the range of the interaction.
For  $n$ separate, non-interacting and indistinguishable interfaces with 
the free energy $F_{\mbox{\tiny{s}}}$ one obtains:
\begin{equation}
\label{zimproved}
 Z_I = \sum_n \;\frac{1}{n!} L_0^n \; \exp(- n F_{\mbox{\tiny{s}}}) = 
     \sum_n \;\frac{1}{n!} \; \exp [- n (F_{\mbox{\tiny{s}}}-\ln L_0) ] \; \; . 
\end{equation}
The sum runs over non-negative even integers in the case of periodic boundary conditions, and positive odd integers in the case of anti-periodic boundary conditions, and the $\frac{1}{n!}$ factor takes into account that the interfaces are indistinguishable. Hence:
\begin{eqnarray}
 \frac{\Za}{\Zp} &=&
 \frac{\sum_{m=0}^{\infty} \;\frac{1}{(2m+1)!}  \; \exp [ - (2m+1) (F_{\mbox{\tiny{s}}}-\ln L_0)]}
      {\sum_{m=0}^{\infty} \;\frac{1}{(2m)!}  \; \exp [ - 2 m (F_{\mbox{\tiny{s}}}-\ln L_0)]} 
\nonumber \\
 &=& \tanh \left\{ \exp [-(F_{\mbox{\tiny{s}}}-\ln L_0) ] \right\} \; \; .
\end{eqnarray}
The solution of this equation with respect to $F_{\mbox{\tiny{s}}}$ provides
us with a second definition of the interface free energy:
\begin{equation}
\label{fs2}
 F_{\mbox{\tiny{s}}}^{(2)} =
 \ln L_0 - \ln \left( \frac{1}{2} \ln \frac{1+\Za/\Zp}{1-\Za/\Zp} \right) \;\;.
\end{equation}
Upon comparison between eq.~(\ref{fs2}) and eq.~(\ref{massandratio}), the tunneling mass $m_{\mbox{\tiny{t}}}$ can be expressed in terms of the interface free energy as:  
\begin{equation}
  m_{\mbox{\tiny{t}}} = 2 \exp(-F_{\mbox{\tiny{s}}}^{(2)}) \; \; ,
\end{equation}
which confirms that the definition $F_{\mbox{\tiny{s}}}^{(2)}$, 
in contrast to $F_{\mbox{\tiny{s}}}^{(1)}$, has a finite, meaningful
$L_0 \rightarrow \infty$ limit. This limit is well approximated
for $L_0 \gg \xi$. All our simulations are done in this regime.
Note that for $L_0 \ll \xi_t$, i.e.  $Z_a/Z_p$ close to zero, 
$F_{\mbox{\tiny{s}}}^{(1)}$ is a good approximation of 
$F_{\mbox{\tiny{s}}}^{(2)}$. For most of our simulations, this condition 
is satisfied.

\section{Interfaces in gauge theory: The effective string perspective}
\label{effectivestringsect}

In quantum gauge theory, the low-energy behaviour of a confined pair of static sources at a distance $r$ might be described by an effective string. In the confining regime, the flux lines between the two sources are squeezed into a thin tube, which might be idealized as a uni-dimensional object. The long-distance properties of the system are dominated by the transverse fluctuations of this tube; 
in this regime, the excitation spectra of the fields in the interior of the tube are expected to be much higher-lying.

Under this assumption, the properties of the system are described through a string partition function, obtained integrating over the possible world-sheet configurations. Each of them has the topology of a cylinder, and contributes a Boltzmann-like factor, whose exponent is given by an effective string action.

In principle, the functional form of the latter is unknown, however it can be constrained, by requiring that it satisfies certain self-consistency properties, and that it yields the correct physical limit for large distances $r$ between the two sources.

This approach underlies the models that have been proposed by Polchinski and Strominger in the 1990's~\cite{Polchinski:1991ax}:
\begin{eqnarray}
\label{polchinskistromingereffectiveaction}
S_{\mbox{\tiny{eff}}} &=& \frac{1}{4\pi} \int d \tau^+ d \tau^- \left[ 
\frac{1}{a^2}(\partial_+ X \cdot \partial_- X ) \right. \nonumber \\ 
& & \left. + \left( \frac{D-26}{12} \right) \frac{ ( \partial_+^2 X \cdot \partial_- X ) (\partial_+ X \cdot \partial_-^2 X )}{(\partial_+ X \cdot \partial_- X )^2} + O(r^{-3})\right] \; ,
\end{eqnarray}
(in which $\tau^{\pm}$ are light-cone world-sheet coordinates, and $a$ is a length scale related to the string tension) and by L\"uscher, Symanzik and Weisz already at the beginning of the 1980's~\cite{Luscher:1980fr,Luscher:1980ac} (see also~\cite{Luscher:2002qv,Luscher:2004ib} for more recent developments):
\eq
\label{serieseff}
S_{\mbox{\tiny{eff}}}= \sigma rL + \mu L + S_0 + S_1 + S_2 + \dots , \;\;\;\;\;\; \mbox{with: } \;\;\;\; S_0=\frac{1}{2} \int d^2 \xi (\partial_a h \partial_a h ) \; ,
\en
(where $L$ denotes the length of the closed world-lines of the static sources, $\sigma$ is the string tension, and $\mu$ is a coefficient associated to a perimeter-like term). The construction of a string action in the form of eq.~(\ref{polchinskistromingereffectiveaction}) allows a generic conformally invariant world-sheet QFT, with the coefficient of the various terms fixed by anomaly cancellation in any dimension $D$. The action can be built by converting the path integral for the collective coordinates of the underlying field theory to covariant form. The $X$ field is unconstrained (and the model still represents a generic interface in $D$ dimensions), but the term appearing in the second line of eq.~(\ref{polchinskistromingereffectiveaction}) takes the Polyakov determinant in conformal gauge into account. Eq.~(\ref{polchinskistromingereffectiveaction}) yields an effective string model which can be expanded around the long-string vacuum. Poincar\'e invariance constrains the $O(r^{-3})$ term of the string spectrum to the form it has in the Nambu-Goto model~\cite{Drummond:2004yp,HariDass:2006sd,Drummond:2006su,Dass:2006ud} (for the definition of the Nambu-Goto model see below). 

In eq.~(\ref{serieseff}), the $S_0$ term describes a conformal model, while the other, higher-dimensional, $S_n$ terms are responsible for the string self-interactions. In this case, the variable $h$ represents a vector with $D-2$ components, that describes the fluctuations transverse to the reference plane (physical gauge). It is interesting to note that the first term beyond $S_0$, which is expected to be a ``boundary term''~\cite{Luscher:2002qv}, is actually forbidden by open-closed string duality~\cite{Luscher:2004ib}.

Among the main implications of the effective string model, we mention the existence of a negative, $O(r^{-1})$ correction to the asymptotic linear potential (the L\"uscher term, which is a Casimir effect), and the logarithmic growth of the square width of the flux tube~\cite{Luscher:1980iy,Caselle:1995fh,Chernodub:2007wi}. Both aspects are related to the fact that, at leading order, the effective string fluctuations in a $D$-dimensional target space can be modelled as $(D-2)$ free, massless bosons.

Since the infinite number of terms appearing on the right-hand sides of eq.~(\ref{polchinskistromingereffectiveaction}) and eq.~(\ref{serieseff}) are not known \emph{a priori}, it is, in general, not possible to work out all-order predictions for the observables of interest. 

Alternatively, one might use an explicit ansatz on the functional form of the effective string action (consistent with the constraints mentioned above, and compatible with the other effective models at its lowest orders): this allows to address the complete mathematical calculation for the expectation values of the physical observables, and to perform an all-order comparison with the numerical results.

A natural choice for the effective string action (for any world-sheet geometry) is the area of the string world-sheet itself. For the case of a closed interface, it can be expressed introducing the $\xi$ coordinates over the interface, and the $g_{\alpha\beta}$ metric induced by the embedding in the target space:
\eq
\label{nambugoto}
S= \sigma \int d^2 \xi \sqrt{\det g_{\alpha\beta}}
\en
(the string tension $\sigma$ has energy dimension $2$).

Eq.~(\ref{nambugoto}) has a natural interpretation in the context of string theory, where it represents the (Euclidean space formulation of) the model due to Nambu and Goto~\cite{Goto:1971ce, Nambu:1974zg}, describing the relativistic quantum dynamics of a purely bosonic string. Although it is well-known that this model is affected by an anomaly (breakdown of rotational symmetry out of the critical space-time dimension $D=26$) and is non-renormalizable (because it is non-polynomial), it has been studied as a possible \emph{effective description} of the low-energy dynamics of confining gauge theories. The reason is that, in the infra-red regime, the lowest-lying degrees of freedom associated with a confining flux tube are transverse fluctuations, and are expected to be modeled by a bosonic string-like dynamics; in that case, $\sigma$ represents the asymptotic value of the string tension. In particular, the geometry of an interface with periodic boundary conditions in both directions would be associated to the description of a torelon, i.e. a string winding around a compact target space, that has been studied in ref.~\cite{Juge:2003vw}. 

A number of implications for this effective description of confinement have been derived theoretically and tested numerically in the literature~\cite{Dass:2005we, Meyer:2004hv, Majumdar:2004qx, Caselle:2004er, Juge:2004xr, Marescathesis, Koma:2003gi, Majumdar:2002mr, Juge:2002br, Kuti:2005xg, Koma:2002rw, Antonovthesis, Caselle:1994df, df83, cp96, Dass:2007tx, HariDass:2006xq, Bringoltz:2006gp, Giudice:2006wz, HariDass:2006pq, Boyko.:2007ae}.

On the other hand, in a condensed matter physics context, eq.~(\ref{nambugoto}) corresponds to the ``capillary wave model''~\cite{Privman:1992zv}; at first order, it describes the transverse fluctuations of a membrane as free, independent, massless modes, whereas the subleading terms introduce (self-)interactions. In this context, $\sigma$ can be interpreted as an (asymptotic) interface tension, which does not depend on the local orientation of the normal to the infinitesimal surface element.

A perturbative expansion in powers of $(\sigma L_1 L_2)^{-1}$ yields the following result for the partition function of the interface with periodic boundary conditions in both directions~\cite{df83,Caselle:1994df,cp96}:
\be
\label{nloequation}
Z=\frac{\lambda}{\sqrt{u}} \exp(-\sigma L_1L_2)
\Bigl| \eta\left(i u\right)/\eta\left(i\right) \Bigr|^{-2}
\left[1+\frac{f(u)}{\sigma L_1L_2}+O\left(\frac{1}{(\sigma L_1L_2)^2}
\right)\right]~~,
\label{2loop}
\ee
where the parameter $\lambda$ can be predicted invoking a perturbative argument for the $\phi^4$ scalar field theory in three dimensions, $\tau=iu=iL_2/L_1$ is the modulus of the torus associated with the cross-section of the system, $\eta$ is Dedekind's function:
\begin{equation}
\eta(\tau)=q^{1/24}\prod_{n=1}^{\infty}\left(1-q^n\right)~~,
\quad\quad q\equiv \exp(2\pi i \tau)~~,
\label{eta}
\end{equation}
and $f(u)$ is defined as: 
\be
f\left(u\right)=\frac{1}{2}\left\{
\left[\frac{\pi}{6} u E_2\left(i u\right)\right]^2 -
\frac{\pi}{6} u E_2\left(i u\right) + \frac{3}{4}\right\}
\label{fu}~~,
\ee
where $E_2(\tau)$ is the first Eisenstein series: 
\begin{equation}
E_2(\tau)=1-24\sum_{n=1}^{\infty}\frac{n~ q^n}{1-q^n}~~.
\quad\quad q\equiv \exp(2\pi i \tau) \; ,
\label{e2}
\end{equation}
In particular, for $u=L_2/L_1=1$ one gets $f(1)=1/4$.

The interface free energy for square lattices of size $L_1=L_2\equiv L$ takes the form:
\eq
F_{\mbox{\tiny{s}}}=\sigma L^2 -\ln\lambda-\frac{1}{4\sigma L^2}+ O\left(\frac{1}{(\sigma L^2)^2}\right) \;\;.
\en
This is the theoretical expectation which, in the following section, will be compared with our numerical results for $F_{\mbox{\tiny{s}}}^{(2)}$ --- see eq.~(\ref{fs2}).

More recently, a different approach to calculate the partition function was proposed in ref.~\cite{Billo:2006zg}: this method is more elegant and powerful with respect to the perturbative expansion in powers of $(\sigma L_1 L_2)^{-1}$, and it takes advantage of the standard covariant quantization techniques for the bosonic string. The power of this method relies in the fact that it allows to resum the complete loop expansion for the interface partition function at all orders, with a final result for the interface partition function in $d$ dimensions $\mathcal{I}^{(d)}$ taking the form of a series of Bessel functions:
\eq
\label{bos21}
\mathcal{I}^{(d)} = 2
\left(\frac{\sigma}{2\pi}\right)^{\frac{d-2}{2}}\, V_T \, \sqrt{\sigma\mathcal{A}u}
\sum_{m=0}^\infty \sum_{k=0}^m c_k c_{m-k} 
\left(\frac{\cale}{u}\right)^{\frac{d-1}{2}}\, K_{\frac{d-1}{2}}\left(\sigma\mathcal{A}\cale\right)~,
\en
(where $\mathcal{A}=L_1 L_2$, and $V_T$ is the product of the system sizes in the transverse directions) and a consistent, closed-form expression for the spectrum levels:
\eq
\label{bos19}
\cale  = \cale_{k,m}  =
\sqrt{1 + \frac{4\pi}{\sigma L_1^2}\left(m-\frac{d-2}{12}\right) + 
\frac{4\pi^{2}}{\sigma^2 L_{1}^{4}} \left( 2k-m \right)^{2}} \;\;,
\en
that agree with those presented in ref.~\cite{Kuti:2005xg}.

For the case of an interface with the boundary conditions of a cylinder, an analogous result was derived in ref.~\cite{Billo:2005iv}, while the associated energy spectrum had already been known since the Eighties~\cite{Arvis:1983fp}.

In eq.~(\ref{bos19}), for $d=3$ and $m=k=0$, the argument of the square root becomes negative for $\sigma L_1^2 < \pi/3$.  This is known as the tachyonic singularity in the effective string framework (see \cite{olesen_dec}) and can be physically  interpreted as the signature of a high temperature deconfinement transition: For temperatures higher than $\Tc/\sqrt{\sigma} = \sqrt{3/\pi}$ the string vanishes and quarks are no longer confined. Equivalently, in the dual model, the Ising spin model, for $L_1<N_t$  the interface tension vanishes and a transition from a ferromagnetic to a paramagnetic phase occurs.

This interpretation was discussed for the first time by Olesen in the case of Polyakov loop correlators~\cite{olesen_dec} and holds essentially unchanged in the present case, although the Arvis spectrum~\cite{Arvis:1983fp} on which Olesen's result was based is very different from the one we have in the present case --- see eq.~(\ref{bos19}). In fact, the lowest state (the one which drives the phase transition) is the same in both spectra.

Notice that one should not expect that this analysis  provides an exact result for the finite temperature transition. Indeed the Svetitsky-Yaffe conjecture~\cite{sy82} suggests that the finite temperature transition in the $\Z_2$ gauge theory belongs to the universality class of the 2D Ising spin model. On the other hand, the analysis of the string picture gives mean-field critical exponents (i.e. $\nu=1/2$). Once more this observation indicates that the Nambu-Goto effective string should be better considered as a mean field description, which is particularly effective at low temperatures and/or large distances.

\section{Method to compute the interface free energy}
\label{methodsect}

Let us define the interface energy as:
\begin{equation}
\Es \equiv \Ea - \Ep \;\;,
\end{equation}
where $\Ea$ ($\Ep$) is the expectation value for the energy of a system with anti-periodic (respectively, periodic) boundary conditions.

The internal energy of a system is given by the derivative of the reduced free energy with respect to $\beta$:
\begin{equation}
\label{energy}
E \equiv - \frac{\partial \ln Z(\beta)}{\partial \beta} =
 \frac{\sum_{\{s\}} \exp [- \beta H(\{s\}) ] \; H(\{s\}) }
      {\sum_{\{s\}} \exp [- \beta H(\{s\}) ] } \;\;.
\end{equation}

From  eq.~(\ref{energy}), by integration from $\beta_0$ to $\beta$ it follows:
\begin{equation}
- \left . \ln \frac{\Za}{\Zp} \right |_{\beta}
= - \left . \ln \frac{\Za}{\Zp} \right |_{\beta_0}
 +  \int_{\beta_0}^{\beta} \mbox{d} \tilde \beta \; \; \Es(\tilde \beta)
\;\;.
\end{equation}
By adding $\ln L_0$ on both sides of the equation we get:
\begin{equation}
\label{masterint}
 F_{\mbox{\tiny{s}}}^{(1)}(\beta)
=F_{\mbox{\tiny{s}}}^{(1)}(\beta_0)
 +  \int_{\beta_0}^{\beta} \mbox{d} \tilde \beta  \; \; \Es(\tilde \beta)
\;\;.
\end{equation}

In general it is difficult to determine free energies directly in a single Monte Carlo simulation. On the other hand, expectation values such as $\Ep$ and $\Ea$ can be easily determined.  

It is rather an old idea to compute free energies (in particular: interface free energies) from eq.~(\ref{masterint})  --- see, e.g., the first few references in~\cite{HaPi97}.  

For $\beta_0$ there are different possible choices: For $\beta < \beta_{\mbox{\tiny{c}}}$ the interface tension vanishes, and with a suitable choice of $\beta_0<\beta_{\mbox{\tiny{c}}}$ (depending on the interface area) $F_{\mbox{\tiny{s}}}^{(1)}(\beta_0) $ vanishes to a very good approximation. Alternatively, one might start the integration from large values of $\beta \gg \beta_{\mbox{\tiny{c}}}$, where $F_{\mbox{\tiny{s}}}^{(1)}(\beta_0)$ can be obtained from the low temperature expansion.

Here we follow the strategy discussed in~\cite{HaPi97}: $F_{\mbox{\tiny{s}}}^{(1)}(\beta_0)$ is computed using the boundary flip algorithm~\cite{Hasenbusch:1992eu} at a $\beta_0$ value corresponding to a $\Za/\Zp$ ratio of the order of $1/10$. For such a choice, $F_{\mbox{\tiny{s}}}^{(1)}(\beta_0)$ can be accurately determined using a moderate amount of CPU time. 

In practice, we have performed  simulations for a finite number of inverse  temperatures $\beta_0 \le \beta_i \le \beta$ to obtain values for $\Es(\beta_i)$. The integration~(\ref{masterint}) must then be performed by using some numerical integration scheme. For a detailed discussion of the schemes that we have used see the appendix~\ref{integrationschemesect}.

In principle, the numerical integration, along with its (small) systematic error, could be avoided. One might compute: 
\begin{eqnarray}
 F(\beta+\Delta \beta) - F(\beta) &=& 
 - \log\left[Z(\beta+\Delta \beta)/Z(\beta)\right] \nonumber \\
&=&- \log \left\{ \langle \exp [ - \Delta \beta H(\{s\}) ] \rangle_{\beta} \right\}
\end{eqnarray}
in a Monte Carlo simulation that generates the Boltzmann distribution corresponding to $\beta$. Here $\Delta \beta$ has to be chosen such that $\Delta \beta \sqrt{\langle H^2 \rangle - \langle H \rangle^2}$ is of order one, to keep the statistical error under control. Then $F(\beta)-F(\beta_0)$ is computed by a sequence of such $\Delta \beta$ steps. 

For an interesting new alternative, using a generalized Jarzynski relation, see ref.~\cite{Chatelain:2007ts}.

Quite a different strategy to compute $F_{\mbox{\tiny{s}}}$ is based on the so-called snake algorithm~\cite{deForcrand:2000fi, Pepe:2001cx}:  A sequence of boundary conditions is introduced, that interpolates between periodic and anti-periodic boundary conditions. The boundary in the 0-direction is progressively filled with $J_{\langle xy \rangle}=-1$. In refs.~\cite{Caselle:2006dv, Billo:2005ej, Caselle:2005vq, Panero:2005iu, Caselle:2005xy, Panero:2004zq, Caselle:2004jq, Caselle:2003db, Caselle:2003rq, Caselle:2002ah, Caselle:2002vq, Caselle:2002rm} we have used this approach to compute ratios of Polyakov-loop correlators. 

In the present work, however, we do not use the snake algorithm, since for boundary conditions close to the anti-periodic boundary conditions the simulations are very difficult. If the last layer of bonds is not completely filled with antiferromagnetic couplings $J_{\langle xy \rangle}=-1$ yet, then it is energetically favorable that the interface sticks to the boundary. On the other hand, there is an entropy gain, when the interface moves freely along the 0-direction. At some stage these two effects are approximately in balance, resulting in extremely long auto-correlation times for any known algorithm.

A major reason for using the numerical integration is that, given the large number of lattice sizes and values of $\beta$, it provides the most efficient way to organize the simulations and to keep the resulting data under control.

\subsection{Monte Carlo simulations}

In order to compute the starting-point free energy $F_{\mbox{\tiny{s}}}(\beta_0)$, we have used a variant of the boundary flip algorithm~\cite{Hasenbusch:1992eu} --- see refs.~\cite{Hasenbusch:1998gh, Caselle:2006dv} for a discussion. The update is performed using the single cluster algorithm~\cite{Wolff:1988uh}. Typically, we have performed  $O(10^7)$ up to $O(10^8)$ measurements. For each measurement, mostly 10 single cluster updates were performed. 

In order to compute the energy for the systems with periodic and anti-periodic boundary conditions, we have used a demonized local Metropolis algorithm implemented in multispin coding technique: details can be found in ref.~\cite{HaPi93}. This way, $n_{\mbox{\tiny{bit}}}$ systems run in parallel in our implementation ($n_{\mbox{\tiny{bit}}}=32$ or $n_{\mbox{\tiny{bit}}}=64$, depending on the machine used). Most simulations were performed with the local algorithm alone.

For each measurement, 12 complete update sweeps were performed. For $n_{\mbox{\tiny{bit}}}=32$ we performed either 100,000 or 96,000 measurements for each copy of the system and each value of $\beta$; for $n_{\mbox{\tiny{bit}}}=64$, 50,000 measurements for each copy of the system were performed. In order to ensure equilibration, we have taken 24,000 local update sweeps. Note that this is an overkill for the smaller lattice sizes and the larger $\beta$ values. However, due to the enormous number of individual simulations, we could not check carefully each of the runs. Therefore we decided to use a common number of thermalization updates, that is suitable for all of the parameter choices we considered --- including the most difficult cases. 

For $\beta$ values close to $\beta_0$, there is a non-negligible probability that more than $0$ (periodic boundary conditions) or $1$ (anti-periodic boundary conditions) interfaces are formed in the system. Most likely, the local update is not capable of generating these additional interfaces within the given number of update cycles. Therefore, we performed single cluster updates in addition. In this step, we could not make use of the multispin coding technique, therefore the cluster update was performed one-by-one for the $n_{\mbox{\tiny{bit}}}$ systems. One single cluster update is performed per measurement; we made no attempt to optimize the ratio of cluster and local updates. For a discussion about a similar combination of algorithms, see ref.~\cite{PlFeLa02}. We have used this method, instead of the local update only, for $\beta=\beta_0+ m \Delta \beta$ with $m \lessapprox 10$.

Going to larger interface areas than those studied in the present work, it would be advisable to use the interface cluster algorithm introduced in ref.~\cite{HM} and further discussed in ref.~\cite{tension}.

In table~\ref{summary_of_runs1} we give a summary of our runs for square interfaces and in table~\ref{summary_of_runs2} for the asymmetric interfaces.  In total, order of thousands individual simulations were performed.

\begin{table}[tbp]
\begin{center}
\begin{tabular}{|r|l|l|l|l|}
\hline
\multicolumn{1}{|c}{$L$} &
\multicolumn{1}{|c}{$\beta_0$} &
\multicolumn{1}{|c}{$\beta_{\mbox{\tiny{max}}}$} &
\multicolumn{1}{|c}{$\Delta \beta$} &
\multicolumn{1}{|c|}{$F_{\mbox{\tiny{s}}}^{(1)}(\beta_0)$} \\
\hline
  8    &   0.23607            &  0.24607      &   0.0002 & 4.69471(36) \\ 
  9    &   0.23407            &  0.24607      &   0.0002 & 4.95168(41) \\ 
 10    &   0.23007            &  0.24607      &   0.0002 & 4.49995(29) \\ 
 11    &   0.23007            &  0.24607      &   0.0002 & 4.98701(37) \\ 
 12    &   0.22907            &  0.24607      &   0.0002 & 5.12142(37) \\ 
 13    &   0.22807            &  0.24607      &   0.0002 & 5.16150(36) \\ 
 14    &   0.22707            &  0.24607      &   0.0002 & 5.10299(34) \\ 
 15    &   0.22667            &  0.24607      &   0.0002 & 5.26780(35) \\ 
 16    &   0.22607            &  0.24607      &   0.0001 & 5.27054(28) \\ 
 17    &   0.225802           &  0.230002     &   0.00005& 5.43001(34) \\ 
 17    &   0.230002          &   0.246202     &   0.0001 &    \\ 
 18    &   0.225302          &   0.230002     &   0.00005& 5.38722(31) \\ 
 18    &   0.230002          &   0.246202     &   0.0001 &    \\ 
 19    &   0.225002          &   0.230002     &   0.00005& 5.44569(32) \\ 
 19    &   0.230002          &   0.246202     &   0.0001 &    \\ 
 20    &   0.224902          &   0.230002     &   0.00005& 5.64529(39) \\ 
 20    &   0.230002          &   0.246202     &   0.0001 &     \\ 
 21    &   0.224702          &   0.230002     &   0.00005& 5.73892(43) \\ 
 22    &   0.224502         &   0.230002     &   0.00005 & 5.80747(41) \\ 
 23    &   0.224302         &   0.230002     &   0.00005 & 5.84791(43) \\ 
 24    &   0.224002         &   0.230002     &   0.00005 & 5.73910(41) \\
 25    &   0.223602         &   0.230002     &   0.00005 & 5.46928(59) \\
 26    &   0.223602        &   0.230002     &   0.00005  & 5.66167(58) \\
 27    &   0.223602        &   0.230002     &   0.00005  & 5.86440(63) \\
 28    &   0.223502    &   0.230002     &   0.00005      & 5.91713(63) \\
 29    &   0.223402    &  0.226102    &   0.00005        & 5.95322(65) \\
 30    &   0.223402 &   0.230002     &   0.00005         & 6.14879(70) \\
 31    &   0.223302 &  0.226102    &   0.00005           & 6.16267(70) \\
 32    &   0.223152  &   0.230002     &   0.00005        & 6.06153(50) \\
 33    &   0.223052  &  0.226102    &   0.00005          & 6.03701(64) \\
 34    &   0.222952  &  0.226102    &   0.00005          & 5.99466(65) \\
 35    &   0.222902  &  0.226102    &   0.00005          & 6.04115(69) \\
 36    &   0.222852    &   0.230002     &   0.00005      & 6.07957(84)  \\
 38    &   0.222752     &  0.226102    &   0.00005       & 6.13242(73) \\
 40    &   0.222652   &   0.230002     &   0.00005       & 6.14808(78) \\
 44    &   0.222552  &   0.230002     &   0.00005        & 6.37175(117) \\
 48    &   0.222452  &   0.230002     &   0.00005        & 6.52109(100) \\
 52    &   0.222352  &    0.226102    &   0.00005        & 6.58323(108)\\
 56    &   0.222252  &    0.226102    &   0.00005        & 6.55408(116) \\
 64    &   0.222152  &    0.226102    &   0.00005        & 6.77107(123) \\
\hline
\end{tabular}
\end{center}
\caption{
Summary of the simulations for the square interfaces. For each linear extension $L=L_1=L_2$ of the interface, the table gives the starting point $\beta_0$ of the integration, the maximal inverse temperature $\beta_{\mbox{\tiny{max}}}$ that has been simulated, and the step-size $\Delta \beta$. In the case of $L=17,18,19,20$ we have two intervals with different $\Delta \beta$. The initial value of the integration $F_{\mbox{\tiny{s}}}^{(1)}(\beta_0)$ has been computed with the boundary flip algorithm. For details see the text.
}
\label{summary_of_runs1}
\end{table}

\begin{table}[tbp]
\begin{center}
\begin{tabular}{|r|l|l|}
\hline
\multicolumn{1}{|c}{$L_1$} &
\multicolumn{1}{|c}{$\beta_0$} &
\multicolumn{1}{|c|}{$F_{\mbox{\tiny{s}}}^{(1)}(\beta_0)$} \\
\hline
24 & 0.222402 & 5.19305(69) \\
28 & 0.222402 & 5.53093(93) \\
32 & 0.222402 & 5.87624(96) \\
36 & 0.222402 & 6.21529(116)\\
40 & 0.222402 & 6.54949(141)\\
44 & 0.222402 & 6.87239(137)\\
48 & 0.222402 & 7.19102(163)\\
\hline
\end{tabular}
\end{center}
\caption{
Summary of runs with asymmetric lattices $L_1 \ne L_2$, in the same notation as in the previous table. $L_0=96$, $L_2=64$, $\Delta \beta=0.00005$ and $\beta_{\mbox{\tiny{max}}}=0.226102$ throughout.
}
\label{summary_of_runs2}
\end{table}

\subsection{Numerical integration}

The numerical evaluation of integral~(\ref{masterint}) was done using standard numerical integration schemes which are summarized in the appendix~\ref{integrationschemesect}.

All the schemes that we have considered can be written in the form:
\begin{equation}
F_{\mbox{\tiny{s}}}^{(1)}(\beta) = F_{\mbox{\tiny{s}}}^{(1)}(\beta_0) + 
 \sum_{j=0}^N  c_j \; \Delta \beta \; \Es(\beta_0 + j \Delta \beta)
                 \;+\;O(N^{-m}) \; \;\;,
\end{equation}
where $\Delta \beta=(\beta-\beta_0)/N$ and $\sum_{j=0}^N c_j = N$. For our final estimates, we have used schemes with an $O(N^{-4})$ integration error. In order to get a quantitative estimate of the integration error, we have compared the result obtained from different schemes; e.g. scheme~(\ref{n4x1}) and scheme~(\ref{n4x2}). 
Furthermore, we have performed the numerical integration for the theoretical predictions of $F_{\mbox{\tiny{s}}}(L_1,L_2,\sigma)$, as discussed in section~\ref{effectivestringsect}, along with $\sigma(\beta)=\sigma_0 (\beta-\beta_{\mbox{\tiny{c}}}) \times [1+ b (\beta-\beta_{\mbox{\tiny{c}}})^\theta +  c (\beta-\beta_{\mbox{\tiny{c}}})]$, with coefficients similar to those reported below. We found that the error of the integration is at least two orders of magnitude smaller than our statistical error, and is hence ignored in the further analysis of the data.

\subsection{Propagation of the statistical error}
\label{crosscorrelation}

The statistical error $\epsilon$ of $F_{\mbox{\tiny{s}}}^{(1)}(\beta)$ is computed using standard error propagation:
\begin{equation}
 \epsilon^2[F_{\mbox{\tiny{s}}}^{(1)}(\beta)] = \epsilon^2[F_{\mbox{\tiny{s}}}^{(1)}(\beta_0)] 
 + (\Delta \beta)^2 \sum_j  c_j^2  \epsilon^2[\Es(\beta_0+j \Delta \beta)]  \;\; ,
\end{equation}
where $\epsilon$ is the statistical error and the $c_j$ coefficient is given by the integration rule.

In order to get correct fits for $F_{\mbox{\tiny{s}}}$ at different values of $\beta$, we have to evaluate the covariances of $F_{\mbox{\tiny{s}}}$ at different values of $\beta$. Let us consider $\beta_2 > \beta_1$: Due to the fact that $F_{\mbox{\tiny{s}}}^{(1)}(\beta_0)$ and $\Es(\tilde \beta)$ with $\tilde \beta\le \beta_1$ are obtained in a common set of simulations, $F_{\mbox{\tiny{s}}}^{(1)}(\beta_1)$ and $F_{\mbox{\tiny{s}}}^{(1)}(\beta_2)$ are statistically correlated.
 
The covariance is defined as: 
\begin{equation}
 \mbox{cov}(A,B) := \langle [A-\langle A \rangle][B-\langle B \rangle] \rangle \;\; .
\end{equation}
In our case: 
\begin{equation}
F_{\mbox{\tiny{s}}}^{(1)}(\beta) = F_{\mbox{\tiny{s}}}^{(1)}(\beta_0)+ \sum_j \tilde{c}_j(\beta) E_j \;\; , 
\end{equation}
where the $E_j$ and $F_{\mbox{\tiny{s}}}^{(1)}(\beta_0)$ are statistically independent. Hence: 
\begin{eqnarray}
\mbox{cov}(F_{\mbox{\tiny{s}}}^{(1)}(\beta_1),F_{\mbox{\tiny{s}}}^{(1)}(\beta_2)) &=&
 \mbox{var}(F_{\mbox{\tiny{s}}}^{(1)}(\beta_0)) + 
  \sum_j \tilde{c}_j(\beta_1) \tilde{c}_j(\beta_2)
  \mbox{var}(E_j) \nonumber \\
&\approx& \mbox{var}(F_{\mbox{\tiny{s}}}^{(1)}(\beta_1)) \;\; .
\end{eqnarray}
The last equality is only approximate, due to the fact that $\tilde{c}_j(\beta_1) \ne 1$ for the last few $j \le m$. In the limit $\Delta \beta \rightarrow 0$, the approximation becomes exact. In our data, we have checked that the approximation is very good, and it is therefore used in the fits.

\section{Numerical results for the interfaces}
\label{interfacesect}

\subsection{Square interfaces}
\label{squareinterfacesubsect}

First we have analysed the data for the square lattices. In comparison to our previous work ref.~\cite{Caselle:2006dv}, we have results for more than four times larger interface areas. This allows us to use the interface tension as a fit parameter, while in ref.~\cite{Caselle:2006dv} we had to take it from ref.~\cite{Caselle:2004jq}, where Polyakov-loop correlators were studied.

Furthermore, here we have data for a large range of inverse temperatures $\beta$, allowing us to address the question of corrections to scaling.

As a starting point, let us first discuss the results for $\beta=0.226102$. Note that for $\beta=0.226102$, the finite temperature phase transition occurs at $L_0=8$~\cite{Caselle:1995wn}.  In table~\ref{FSNT8} we have summarized our results for the interface free energy $F_{\mbox{\tiny{s}}}^{(2)}$ at $\beta=0.226102$. Note that here we have converted, using eqs.~(\ref{fs1},\ref{fs2}), $F_{\mbox{\tiny{s}}}^{(1)}$, which is the result of our numerical integration, to $F_{\mbox{\tiny{s}}}^{(2)}$, which is less dependent on $L_0$ and is therefore more suitable for the comparison with the theoretical predictions. For $L_0/L$ fixed, the difference between $F_{\mbox{\tiny{s}}}^{(2)}$ and $F_{\mbox{\tiny{s}}}^{(1)}$ goes down exponentially as the interface area increases. For our numerical results at $\beta=0.226102$ this difference is larger than the statistical error only for $L\le 20$.

\begin{table}[tbp]
\begin{center}
\begin{tabular}{|c|l|}
\hline
\multicolumn{1}{|c}{$L$} &
\multicolumn{1}{|c|}{$F_{\mbox{\tiny{s}}}^{(2)}$} \\
\hline
   18  &  \phantom{0}6.00956(34) \\
   19  &  \phantom{0}6.40442(38) \\ 
   20  &  \phantom{0}6.81999(45) \\ 
   21  &  \phantom{0}7.25617(50) \\ 
   22  &  \phantom{0}7.71334(51) \\ 
   23  &  \phantom{0}8.19024(56) \\  
   24  &  \phantom{0}8.68757(58) \\ 
   25  &  \phantom{0}9.20809(77) \\
   26  &  \phantom{0}9.74659(78) \\
   27  &  10.30706(84) \\ 
   28  &  10.88919(87) \\
   29  &  11.48975(92) \\
   30  &  12.11320(98) \\
   31  &  12.7558(10) \\
   32  &  13.4210(11) \\
   33  &  14.1074(12) \\  
   34  &  14.8145(13) \\
   35  &  15.5415(13) \\
   36  &  16.2881(15) \\
   38  &  17.8477(16) \\ 
   40  &  19.4919(17) \\  
   44  &  23.0292(22) \\
   48  &  26.9095(24) \\ 
   52  &  31.1193(28) \\
   56  &  35.6618(32) \\
   64  &  45.7769(40) \\
\hline
\end{tabular}
\end{center}
\caption{
Interface tension $F_{\mbox{\tiny{s}}}^{(2)}$ for square interfaces at $\beta=0.226102$ as a function of the linear lattice size $L$. 
}
\label{FSNT8}
\end{table}

Similarly to ref.~\cite{Caselle:2006dv} we have fitted the data with the ans\"atze:
\begin{equation}
\label{ansatz1}
 F_{\mbox{\tiny{s}}}^{(2)} = \sigma L^2 + c_0 + \frac{c_2}{\sigma L^2}
\end{equation}
and
\begin{equation}
\label{ansatz2}
 F_{\mbox{\tiny{s}}}^{(2)} = \sigma L^2 + c_0 + \frac{c_2}{\sigma L^2} + 
             \frac{c_4}{(\sigma L^2)^2} \;\;,
\end{equation}
where $\sigma$, $c_0$, $c_2$ and $c_4$ are the free parameters of the fits. At this stage of the analysis we made no attempt to compare with the full NG-prediction which can be obtained from eq.~(\ref{bos21}). 

Results of fits with the ansatz (\ref{ansatz1}) are given in table~\ref{FIT1NT8}. Starting from $L_{\mbox{\tiny{min}}}=19$, the $\chi^2/$d.o.f. is smaller than one. The fit result for $c_2$ is, up to $L_{\mbox{\tiny{min}}}=24$, decreasing with increasing $L_{\mbox{\tiny{min}}}$. For $L_{\mbox{\tiny{min}}}=24$ we get $c_2=-0.246(13)$ which is consistent with the NG prediction $c_2=-0.25$. Since the fit result is increasing with $L_{\mbox{\tiny{min}}}$ we might consider $c_2=-0.246(13)$ as an upper bound.

\begin{table}[tbp]
\begin{center}
\begin{tabular}{|c|l|l|l|c|}
\hline
\multicolumn{1}{|c}{$L_{\mbox{\tiny{min}}}$} &
\multicolumn{1}{|c}{$\sigma$} &
\multicolumn{1}{|c}{$c_0$} &
\multicolumn{1}{|c}{$c_2$} &
\multicolumn{1}{|c|}{$\chi^2/$d.o.f.} \\
\hline
18 &  0.0105283(7) &  2.6589(11) & -0.209(4)  & 1.40 \\
19 &  0.0105273(8) &  2.6611(13) & -0.219(5)  & 0.95 \\
20 &  0.0105267(8) &  2.6625(15) & -0.226(6)  & 0.81 \\
21 &  0.0105264(9) &  2.6635(17) & -0.230(7)  & 0.79 \\
22 &  0.0105262(9) &  2.6640(19) & -0.234(9)  & 0.82 \\
23 &  0.0105257(10)&  2.6654(22) & -0.242(11) & 0.76 \\
24 &  0.0105255(11)&  2.6661(25) & -0.246(13) & 0.78 \\
25 &  0.0105262(12)&  2.6636(30) & -0.229(17) & 0.69 \\
26 &  0.0105259(13)&  2.6649(34) & -0.239(21) & 0.70 \\
\hline
\end{tabular}
\end{center}
\caption{
Fits with ansatz~(\ref{ansatz1}) for square interfaces at $\beta=0.226102$.
}
\label{FIT1NT8}
\end{table}

As a check we have performed fits with the ansatz~(\ref{ansatz2}); the results are summarized in table \ref{FIT2NT8}. The $\chi^2/$d.o.f. is below one for all $L_{\mbox{\tiny{min}}}$ available.  $c_2$ is now increasing with increasing $L_{\mbox{\tiny{min}}}$. Unfortunately, no stable estimate for $c_4$ is obtained. Higher order corrections seem to play an important r\^ole. The results $c_2 = -0.280(38)$ from $L_{\mbox{\tiny{min}}}=20$ might serve as lower bound for $c_2$. Combining the results of the fits with the ans\"atze~(\ref{ansatz1},\ref{ansatz2}) we might summarize our results as: $-0.246(13)  >  c_2 > -0.280(38) $, which is fully consistent with the theoretical prediction. As our final result for the interface tension we take $\sigma(0.226102)=0.0105255(11)$, obtained from the ansatz~(\ref{ansatz1}) and $L_{\mbox{\tiny{min}}}=24$.  The comparison with results from the ansatz~(\ref{ansatz1}) suggests that systematic errors should not be larger than the statistical error that is quoted.

We have repeated this type of analysis for $\beta=0.223102$, $0.223452$, $0.223952$, $0.224752$, $0.227202$, $0.228802$, $0.230002$, $0.236025$, $0.24$ and $0.24607$. Throughout we find for fits with the ansatz~(\ref{ansatz1}) that the numerical result for $c_2$ is decreasing with increasing $L_{\mbox{\tiny{min}}}$, until it starts to fluctuate. In the case of the fits with ansatz~(\ref{ansatz2}) we see that $c_2$ is increasing with increasing $L_{\mbox{\tiny{min}}}$ for $\beta<0.230002$; for larger values of $\beta$ it decreases. For $\beta<0.230002$, our results are consistent with the theoretical prediction (for the scaling limit) $c_2=-0.25$. For larger values of $\beta$, deviations become visible: for instance, for $\beta=0.24$, the result from a fit with ansatz~(\ref{ansatz1}) and $L_{\mbox{\tiny{min}}}=12$ is $c_2=-0.31(2)$. 

\begin{figure}
\includegraphics[height=10cm]{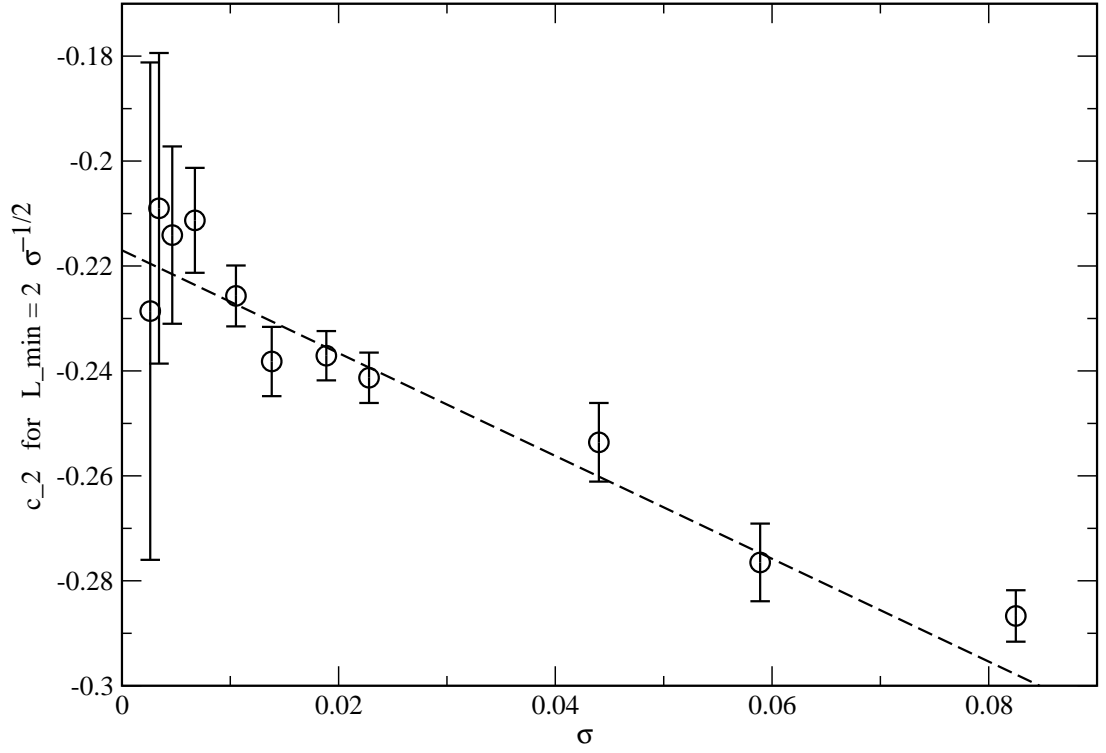}
\vskip0.5cm
\caption{Results for $c_2$ from fits with the ansatz~(\ref{ansatz1})
and $L_{\mbox{\tiny{min}}} = 2/\sqrt{\sigma}$. 
\label{figc2corr}
}
\end{figure}

In order to disentangle corrections to scaling from truncation effects in the ansatz, we have studied $c_2$ as obtained from fits with the ansatz~(\ref{ansatz1}) and $\sqrt{\sigma} L_{\mbox{\tiny{min}}} \approx 2$ fixed. Na\"{\i}vely fitting all data for $\beta \le 0.24$ we get: 
\begin{equation}
c_2|_{L_{\mbox{\tiny{min}}}=2/\sqrt{\sigma}} = -0.217(4) -0.98(15) \sigma~~,
\end{equation}
with $\chi^2/$d.o.f.$=0.55$. In figure~\ref{figc2corr} we show our data along with the result of this fit. We conclude that $c_2$ is affected by  corrections to scaling. However the corrections are, within the numerical precision,  proportional 
to $\sigma \propto \xi^{-2}$, i.e. they vanish much faster than $\xi^{-\omega}$, where $\omega=0.821(5)$~\cite{DB03} is the exponent of the leading correction to scaling.\footnote{Field theoretical methods give slightly smaller values: $\omega=0.814(18)$ from  the  $\epsilon$-expansion and $\omega=0.799(11)$ from perturbation theory in three dimensions fixed~\cite{GuidaZinn}. In ref.~\cite{Hasenbusch:1999mw} the  value $\omega=0.845(10)$ was obtained from Monte Carlo simulations of the $\phi^4$ model on the lattice.} This could be explained by the fact that the effective interface model only assumes  that the symmetries of the continuous space are restored. The restoration of rotational symmetry is indeed  associated with a correction exponent $\omega' \approx 2$~\cite{pisa97}.

\begin{table}[tbp]
\begin{center}
\begin{tabular}{|c|l|l|l|l|c|}
\hline
\multicolumn{1}{|c}{$L_{\mbox{\tiny{min}}}$} &
\multicolumn{1}{|c}{$\sigma$} &
\multicolumn{1}{|c}{$c_0$} &
\multicolumn{1}{|c}{$c_2$} &
\multicolumn{1}{|c}{$c_4$} &
\multicolumn{1}{|c|}{$\chi^2/$d.o.f.} \\
\hline
18 & 0.0105246(11) &  2.6701(29) & -0.299(22) & 0.20(5) &  0.71 \\
19 & 0.0105250(12) &  2.6689(35) & -0.288(29) & 0.17(7) &  0.72 \\
20 & 0.0105252(14) &  2.6682(42) & -0.280(38) & 0.14(10)&  0.75 \\
21 & 0.0105252(15) &  2.6678(51) & -0.275(50) & 0.13(15)&  0.79 \\
22 & 0.0105252(15) &  2.6679(56) & -0.276(60) & 0.14(19)&  0.84 \\
\hline
\end{tabular}
\end{center}
\caption{
Fits with ansatz~(\ref{ansatz2}) for square interfaces at $\beta=0.226102$.
}
\label{FIT2NT8}
\end{table}

\begin{figure}
\includegraphics[height=10cm]{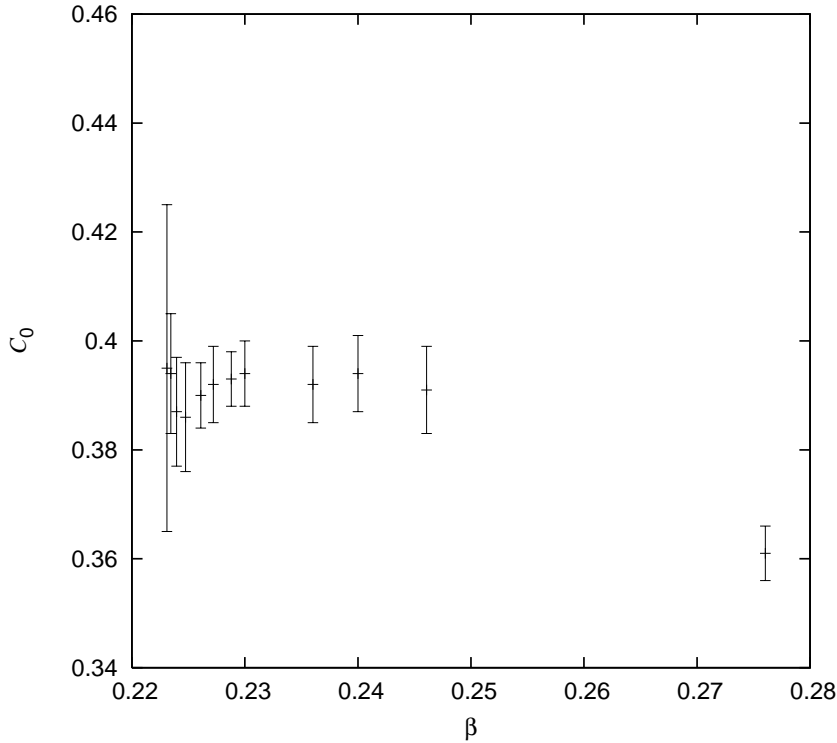}
\vskip0.5cm
\caption{The constant $C_0$ as a function of $\beta$. The quantity is defined in eq.~(\ref{defC0}) in the text. The value for $\beta=0.27604$ is taken from ref.~\cite{Caselle:2006dv}. 
\label{finalC0plot}
}
\end{figure}

Next we study the behaviour of $c_0$. In the scaling limit, this quantity should behave like: 
\begin{equation}
\label{defC0}
 c_0(\beta) = C_0 -\frac{1}{2} \ln [ \sigma(\beta) ] \;.
\end{equation}
In fig.~\ref{finalC0plot} we have plotted our results for $C_0$ as a function
of $\beta$. 

Within our numerical precision there is no sign of corrections to scaling whatsoever
for $\beta\le 0.24607$. Only for $\beta=0.27604$, the value is taken from table 5 of ref.~\cite{Caselle:2006dv}, a clear deviation is visible. Unfortunately, we have no clear theoretical understanding why corrections to scaling should be so small in this quantity.

Results for the interface tension are summarized in table \ref{finalsigma}. All these results are taken from fits with the ansatz~(\ref{ansatz1}) using $L_{\mbox{\tiny{min}}} \approx 2.5/\sqrt{\sigma}$. As discussed above, for the case $\beta=0.221602$, systematic errors should be smaller than the statistical error that is quoted. 

\subsection{Global fit of the data for the interface free energy}
\label{sect:5.1}
In the neighbourhood of the transition, the interface tension behaves as:
\begin{equation}
\label{powerlaw}
\sigma(\beta) = \sigma_0 t^{\mu} \times (1 + a t^{\theta} + b t  
+ c t^{2 \theta} + d t^{\theta'} + ...) \;\;,
\end{equation}
where $t=\beta-\beta_{\mbox{\tiny{c}}}$ is the reduced temperature. The most accurate result for the inverse critical temperature is $\beta_{\mbox{\tiny{c}}}=0.22165455(5)$~\cite{DB03}. The critical exponent $\mu$ of the interface tension is related with the critical exponent of the 
correlation length as $\mu=2 \nu$. The most accurate values given in the literature are $\nu=0.63012(16)$~\cite{phi4pisa} from the analysis of high temperature series expansions, and $\nu=0.63020(12)$~\cite{DB03} from a finite size scaling analysis of Monte Carlo data. The same study provides $\omega=0.821(5)$, hence $\theta=\nu \omega=0.5174(33)$. Note that the value of $\theta'=1.05(7)$~\cite{NeRi84} has quite a large uncertainty. 

Since $2 \theta \approx \theta' \approx 1$, we take as ansatz for the fits:
\begin{equation}
\label{powerfit}
\sigma(\beta) = \sigma_0 t^{\mu} \times (1 + a t^{\theta} + b t) \;.
\end{equation}
Fitting our new results for $\sigma(\beta)$ to this ansatz would require to take into account the cross-correlations among the values of $\sigma(\beta)$ for different values of $\beta$. Instead of computing these cross-correlations, we performed fits for $F_{\mbox{\tiny{s}}}^{(2)}$ fitting the $L$ and the $\beta$ dependence at the same time. The cross-correlations of $F_{\mbox{\tiny{s}}}^{(2)}$ at different values of $\beta$ can be easily obtained as discussed in subsection~\ref{crosscorrelation}.

Based on the results obtained above, we performed a four parameter fit of the data for $F_{\mbox{\tiny{s}}}(L,\beta)$. To this end, we have used the ansatz:
\begin{equation}
\label{finiteansatz}
 F_{\mbox{\tiny{s}}}(L,\beta) = \sigma(\beta) L^2 + C_0 -\frac{1}{2} \ln[\sigma(\beta)]
              -\frac{1}{4} \frac{1}{\sigma(\beta) L^2} \; ,
\end{equation}
where the interface tension is given by the ansatz~(\ref{powerfit}), namely, the free parameters are $\sigma_0$, $a$, $b$ and $C_0$. The critical exponents and the inverse critical temperature are fixed by their best estimates given in the literature, which are quoted above eq.~(\ref{powerfit}).

In the fit, we have used results for the interface free energy at the same values of $\beta$ as discussed in the previous subsection: $\beta=0.223102$, $0.223452$,
$0.223952$, $0.224752$, $0.227202$, $0.228802$, $0.230002$, $0.236025$, $0.24$ and $0.24607$. Our data would allow to use more values of $\beta$. However, little information would be added this way, since, by construction, the interface free energies at close-by values of $\beta$ are highly correlated.

After some experimenting we decided to take our final estimate from a fit with input data characterized by $\beta \le 0.227202$, and $F_{\mbox{\tiny{s}}}-\ln L \ge 8$, which roughly corresponds to $\sqrt{\sigma} L \ge 3$. In total, 51 data-points  satisfy this criterion. The results for the fit parameters are $\sigma_0=10.083(8)$, $a=-0.479(26)$, $b=-2.12(19)$ and $C_0=0.3895(8)$, where $\chi^2$/d.o.f. $=0.79$. 

In order to check the $L$ dependence of our result, we have repeated the fit with $\beta \le 0.227202$, and $F_{\mbox{\tiny{s}}}-\ln L \ge 4$ (which corresponds roughly to $\sqrt{\sigma} L \ge 2.1$)  and $L \le L_{\mbox{\tiny{max}}}=44$.  This means that the range in $L$ is roughly $\sqrt{2}$ times smaller than that of the previous fit. In total, 84 data points satisfy this criterion. The results of this fit are $\sigma_0=10.080(8)$, $a=-0.471(24)$, $b=-2.20(17)$ and $C_0=0.3915(4)$ with $\chi^2$/d.o.f. $=0.89$.

Next, we changed the $\beta$-interval of our fit. We included data that satisfy the criteria $0.224302 \le \beta \le 0.233$, $F_{\mbox{\tiny{s}}}-\ln L \ge 8$, $L \le L_{\mbox{\tiny{max}}}=44$. There are 55 data points that satisfy these criteria. The results of the fit are $\sigma_0=10.085(5)$, $a=-0.484(12)$, $b=-2.08(7)$ and $C_0=0.3886(6)$ with $\chi^2$/d.o.f. $=0.64$.

The results for $\sigma_0$ are consistent among the three different fits. The differences, possible within the statistical error, of these results provide an estimate of the systematic error due to finite-$L$ and large-$t$ corrections that are not taken into account by the ansatz. We arrive at:
\begin{eqnarray}
\label{sigmafinal}
\sigma_0=10.083(8)[26] + \phantom{xxxxxxxxxxxxxxxxxxxxxxxxxxxxxxxxxxx} \\
 22330  (\beta_{\mbox{\tiny{c}}} - 0.22165455) + 
         174  (\nu -0.6302)   - 0.237  (\theta - 0.5174)
\nonumber
\end{eqnarray}
and 
\begin{eqnarray}
\label{afinal}
a= -0.479(26)[120]  \phantom{xxxxxxxxxxxxxxxxxxxxxxxxxxxxxxxxxxxxx} \\
 - 55866  (\beta_{\mbox{\tiny{c}}} - 0.22165455) 
       -149   (\nu -0.6302)  - 2.82  (\theta - 0.5174) \; .
\nonumber
\end{eqnarray}
The number in the brackets gives the systematic error caused by corrections to the ansatz, as discussed above.  In the second line of eqs.~(\ref{sigmafinal},\ref{afinal}) we give the dependence of the result on the input parameters $\beta_{\mbox{\tiny{c}}}$, $\nu$ and $\theta$. The dependence of $C_0$ on $\beta_{\mbox{\tiny{c}}}$, $\nu$ and $\theta$ is small enough to be neglected.  We take: 
\begin{equation}
C_0=0.3895(8)
\end{equation}
as our final result. The comparison of the three fits done above suggests that the systematic should not be larger than the statistical error. In table~\ref{finalsigma} we compare results for $\sigma$ obtained from the global fit with the results obtained from  analysing single values of $\beta$ in the previous subsection. For all values of $\beta$ the results are consistent.

\subsection{Comparison with the literature}

Using the definitions $\tilde t = (\beta-\beta_{\mbox{\tiny{c}}})/\beta_{\mbox{\tiny{c}}}$  and $\sigma = \tilde \sigma_0 \tilde t^{\mu} \times (1 + ...)$ we get: 
\begin{equation}
\tilde \sigma_0 =  1.510(4) \; ,
\end{equation}
where the error is dominated by the uncertainty of $\nu$. Note that this result is perfectly consistent with (and more precise than) the most accurate result $\tilde \sigma_0 = 1.50(1)$ given in the literature~\cite{ZinnFisher} using Monte Carlo data of ref.~\cite{HaPi93}. A more comprehensive list of results for $\tilde \sigma_0$ is given in table 8 of ref.~\cite{tension}.

\begin{table}[tbp]
\begin{center}
\begin{tabular}{|l|l|l|}
\hline
\multicolumn{1}{|c}{$\beta$} &
\multicolumn{1}{|c}{$\sigma$ global fit} &
\multicolumn{1}{|c|}{$\sigma$} \\
\hline
0.223102 &  0.0026083(6)(7)  & 0.0026043(53) \\
0.223452 &  0.0034176(6)(8)  & 0.0034152(31) \\
0.223952 &  0.0046397(6)(11) & 0.0046384(26) \\  
0.224752 &  0.0067258(6)(16) & 0.0067269(17) \\
0.226102 &  0.0105254(7)(28) & 0.0105255(11) \\  
0.227202 &  0.0138217(8)(42) & 0.0138220(17) \\
0.228802 &                   & 0.0188659(13) \\ 
0.230002 &                   & 0.0228068(12) \\ 
0.233    &                   & 0.033114(15) \\
0.236025 &                   & 0.044019(9)  \\
0.24     &                   & 0.058913(5)  \\
0.24607  &                   & 0.082510(5) \\
\hline
\end{tabular}
\end{center}
\caption{
Final results for $\sigma$ at given values of the inverse temperature $\beta$. In the second column we give $\sigma$ as obtained from our global fit. The statistical error of $\sigma$ is properly computed, and the second error quoted is the systematic one. It is estimated from comparing results from different fit ranges. The third column gives  $\sigma$ obtained from fits with the ansatz (\ref{ansatz1}) and $L_{\mbox{\tiny{min}}} \approx 2.5/\sqrt{\sigma}$. For the three smallest values of $\beta$ the global fit provides more accurate results for $\sigma$ than the fits with the ansatz (\ref{ansatz1}). 
}
\label{finalsigma}
\end{table}

Also our results for the interface tension at given values of $\beta$ can be compared with values given in the literature. Here we only give a small selection of the most recent results. For more see e.g. ref.~\cite{wholevolume}. In~\cite{Caselle:2004jq}, studying Polyakov-loop correlators,  we find $\sigma=0.0105241(15)$ and $0.044023(3)$ for $\beta=0.226102$  and $0.236025$, respectively. These values are completely consistent with the estimates of the present work. One should note that they are obtained using a completely different
numerical procedure.

In Fig. 4 of ref.~\cite{Juge:2004xr} the results $\sigma=0.004782(6)$, $0.01011(10)$, $0.022798(2)$ and $0.02752(10)$ for $\beta=0.224$, $0.226$, $0.23$ and $0.23142$ are provided. \\
\noindent
These can be compared with $\sigma=$ $0.004761(2)$, $0.010228(4)$, $0.022800(1)$ and $0.027603(2)$ for the same values of $\beta$, taken from our global fit. Our values are  consistent with those of ref.~\cite{Juge:2004xr}, except for $\beta=0.224$, where we observe a discrepancy by three and a half standard deviations.

In ref.~\cite{tension} the interface tension has been computed in a similar way as in the present work. Our present results are by approximately a factor of ten more precise than those of ref.~\cite{tension}. The results quoted in ref.~\cite{tension} are consistent with our present estimates within two standard deviations.

The results of ref.~\cite{Caselle:1994df}, using the two-loop approximation to fit the data, are $\sigma=0.004778(14)$, $0.006547(69)$, $0.009418(61)$ and $0.014728(40)$ for $\beta=0.2240$, $0.2246$, $0.2258$ and $0.2275$ to be compared with $\sigma=0.004761(2)$, $0.006319(2)$, $0.009418(61)$ and $0.014740(5)$. While the results for $\beta=0.2240$ and $0.2275$ are perfectly consistent, there is a mismatch by 3.3 and 3.8 standard deviations in the case of $\beta=0.2246$ and $0.2258$, respectively. This is likely due to the fact that for these two values of $\beta$ only small interface areas were available and too small areas had been included into the fit.

\subsection{Square interfaces: small $L$}

\begin{figure}
\includegraphics[height=10cm]{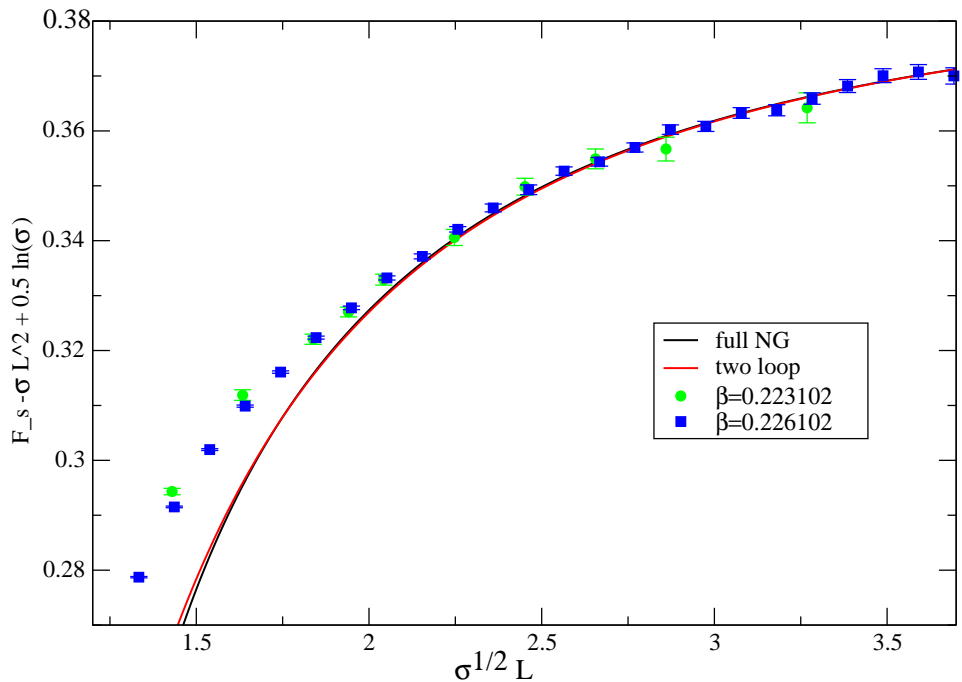}
\vskip0.5cm
\caption{Comparison of the 2-loop prediction, the full Nambu-Goto result with the data for square interfaces at $\beta=0.223102$ and $\beta=0.226102$.
\label{small2}
}
\end{figure}

In figure \ref{small2} we plot $F_{\mbox{\tiny{s}}}^{(2)}-\sigma L^2 + 0.5 \ln(\sigma)$ as a function of $\sqrt{\sigma} L$. The numerical values for $\sigma$ are taken 
from table \ref{finalsigma}. Note that these values of $\sigma$ are obtained from rather large values of $\sqrt{\sigma} L$ (i.e. are little affected by higher order corrections). In addition to the numerical data for square interfaces at $\beta=0.223102$ and $\beta=0.226102$ (roughly corresponding to the critical values of $\beta$ for which the finite temperature transition occurs for $N_t = 16$ and $N_t = 8$, where $N_t$ denotes the number of lattice spacings in the ``inverse temperature'' compactified direction) we give the 2-loop prediction and the full Nambu-Goto result. In the case of the string predictions we have taken $C_0=0.3895$ into account. We observe that, numerically, for $\sqrt{\sigma} L \gtrapprox  1.6$ 
there is very little difference between the two-loop approximation and the full NG result. By expanding eq.~(\ref{bos21}) we find that the coefficient of the $1/(\sigma L^2)^2$ term for the full NG result is approximately equal to $-0.017$. Within the statistical error, the Monte Carlo results for the two values of $\beta$ fall on top of the  2-loop and full NG predictions for $\sqrt{\sigma} L \gtrapprox 2.2$; note that the 1-loop approximation predicts $F_{\mbox{\tiny{s}}}-\sigma L^2$ to be constant. For smaller $\sqrt{\sigma} L$, the data rather abruptly depart from the string prediction. This indicates that (mainly) not $O(1/(\sigma L^2)^2)$ but rather higher order corrections are responsible for the deviation. Still down to $\sqrt{\sigma} L \approx 1.8$ the Monte Carlo data for the two values of $\beta$ fall on top of each other within the error-bars. For smaller $\sqrt{\sigma} L$, differences become visible, indicating corrections to scaling. These scales should be compared with the scale of the finite temperature transition $\sqrt{\sigma} N_t \approx 0.81$ (for a discussion of this number see section \ref{amplituderatiosect} below; the effective string model gives $\sqrt{\sigma} N_t = \sqrt{\pi/3} \approx 1.023$).
 
\subsection{Asymmetric interfaces}
\label{asymmetricinterfacesubsection}

\begin{figure}
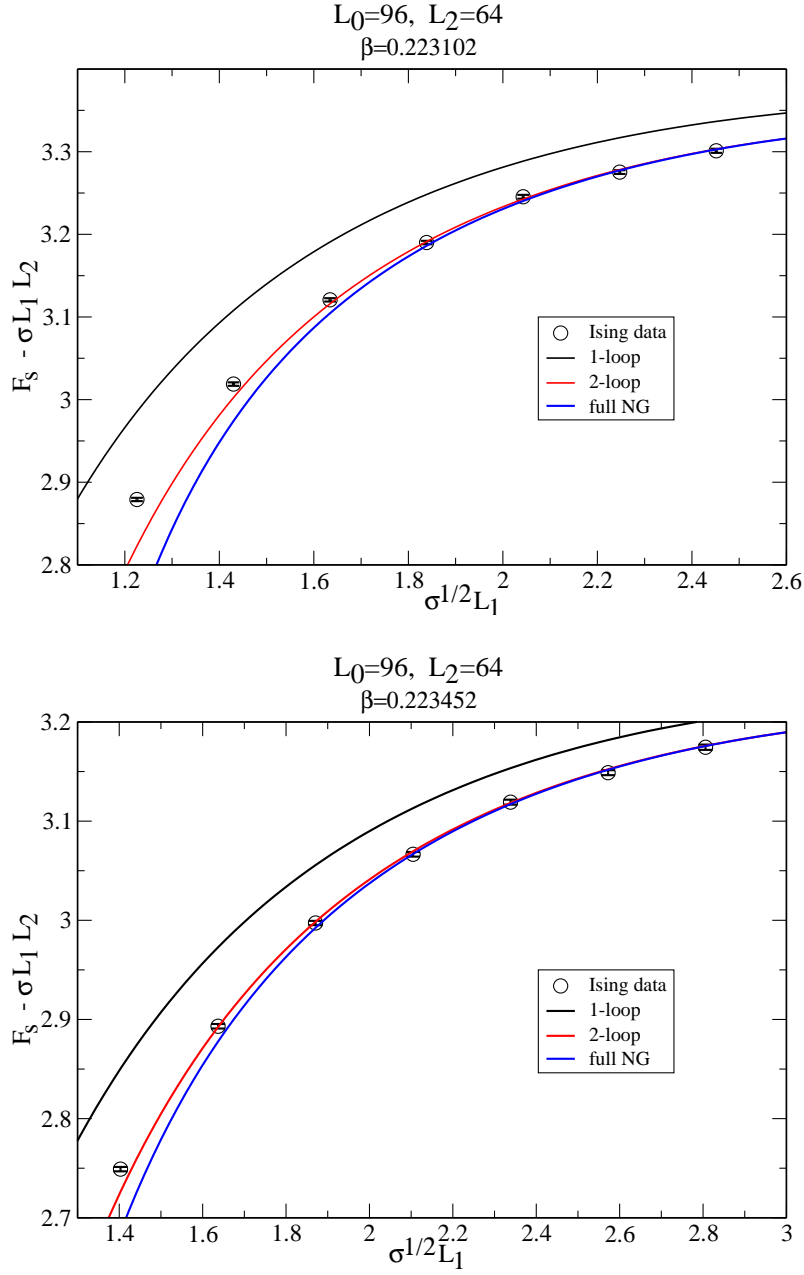

\includegraphics[height=8.1cm]{final64b0.223102.eps}
\vskip0.55cm
\includegraphics[height=8.1cm]{final64b0.223452.eps}
\vskip0.2cm
\caption{
In the two figures we give $F_{\mbox{\tiny{s}}}^{(2)}- \sigma L_1 L_2$ as a function of $\sqrt{\sigma} L_1$ for $\beta=0.223102$ and $\beta=0.223452$. In  all cases $L_0=96$ and $L_2=64$. Note that in the case of the Monte Carlo results the statistical error is smaller than the symbol (circle). The 1-loop, 2-loop and full NG predictions are given as solid black, red and blue lines, respectively. 
\label{asymmetric0223102F}
}
\end{figure}

\begin{figure}
\includegraphics[height=8.5cm]{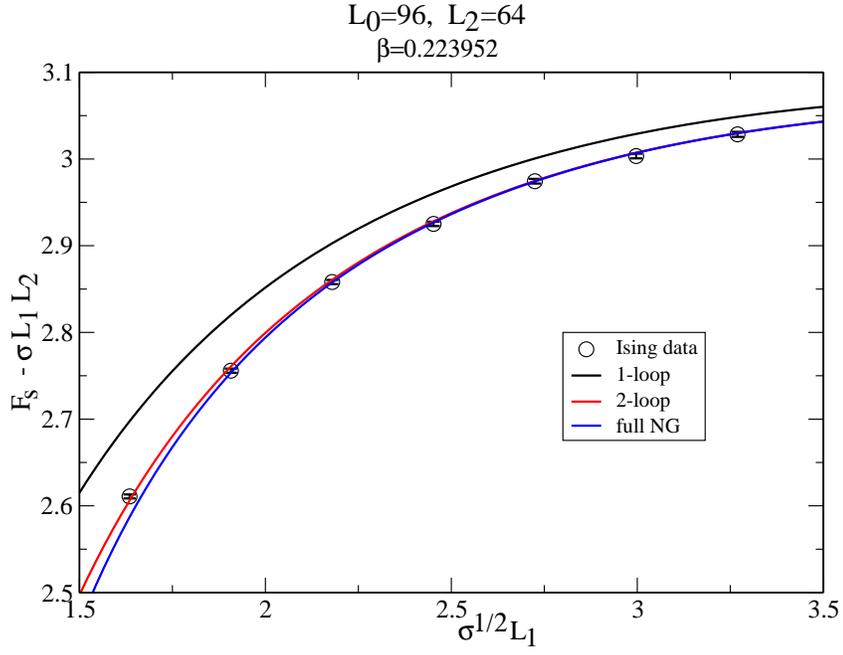}
\vskip0.7cm
\includegraphics[height=8.5cm]{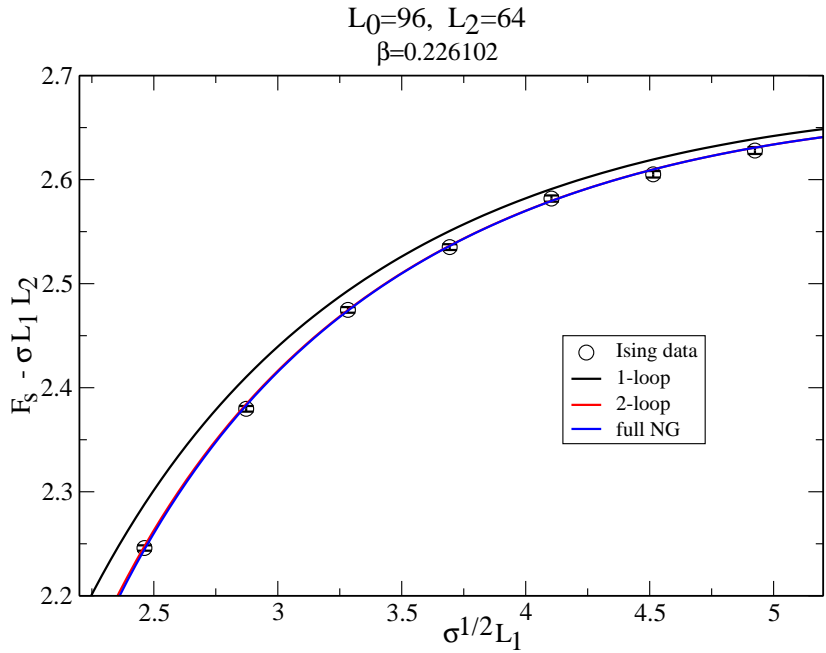}
\vskip0.2cm
\caption{
Same as in the previous figure, but for $\beta=0.223952$ and $\beta=0.226102$. For $\beta=0.226102$, the 2-loop and full NG predictions fall (within the resolution of our plot) on top of each other.
\label{asymmetric226102F}
}
\end{figure}

In this subsection we compare our results for rectangular interfaces with the effective string predictions discussed in sect.~\ref{effectivestringsect}. Note that the effective string results have quite a non-trivial dependence on the aspect ratio $u=L_2/L_1$. Therefore this comparison is rather a stringent test of the theoretical predictions. We have simulated lattices with the linear sizes $L_0=96$ and $L_2=64$. In the remaining direction, the linear size assumes the values $L_1=24,28,32,36,40$ or $48$. We use the values of $\sigma$ and $C_0$ obtained in the section~\ref{squareinterfacesubsect} as input for the theoretical predictions. Therefore we have no free parameters to fit, and we can perform a direct comparison among simulations and theoretical predictions. We have computed the interface free energy $F_{\mbox{\tiny{s}}}^{(2)}$  at $\beta=0.223102$, $0.223452$, $0.223952$ and $0.226102$, corresponding to $N_t=16$, $14$, $12$ and $8$. Note that only for $\beta=0.223102$ and $L_1=24$ the difference between $F_{\mbox{\tiny{s}}}^{(2)}$ and $F_{\mbox{\tiny{s}}}^{(1)}$ is slightly larger than the statistical error. Our numerical results, along with the various string predictions are summarized in tables~\ref{asymmetric0223102} to~\ref{asymmetric0226102}, and shown in the plots~\ref{asymmetric0223102F} and~\ref{asymmetric226102F}.

\begin{table}[tbp]
\begin{center}
\begin{tabular}{|c|c|c|c|c|c|c|c|}
\hline
$L_{1}$ & $L_{2}$ & $L_{3}$ & $F_{\mbox{\tiny{s}}}^{(1) \mbox{\tiny{num}}}(\beta)$ & $F_{\mbox{\tiny{s}}}^{(2) \mbox{\tiny{num}}}(\beta)$ & $F_{\mbox{\tiny{s}}}^{\mbox{\tiny{th}}}(\beta)$, full & $F_{\mbox{\tiny{s}}}^{\mbox{\tiny{th}}}(\beta)$, 1-l. & $F_{\mbox{\tiny{s}}}^{\mbox{\tiny{th}}}(\beta)$, 2-l. \\
\hline
   24 & 64 &  96 &  6.8887(20)  & 6.8855(20)  &  6.7495 &  6.9974 &  6.8347 \\ 
   28 & 64 &  96 &  7.6935(21)  & 7.6929(21)  &  7.6537 &  7.7875 &  7.6821 \\ 
   32 & 64 &  96 &  8.4626(20)  & 8.4625(20)  &  8.4518 &  8.5380 &  8.4632 \\ 
   36 & 64 &  96 &  9.1996(21)  &             &  9.2012 &  9.2632 &  9.2062 \\ 
   40 & 64 &  96 &  9.9227(23)  &             &  9.9231 &  9.9713 &  9.9253 \\ 
   44 & 64 &  96 & 10.6203(23)  &             & 10.6278 & 10.6674 & 10.6288 \\ 
   48 & 64 &  96 & 11.3138(25)  &             & 11.3209 & 11.3548 & 11.3213 \\ 
\hline
\end{tabular}
\end{center}
\caption{Free energy of rectangular interfaces, at $\beta=0.223102$. The value of $F_{\mbox{\tiny{s}}}^{(2) \mbox{\tiny{num}}}(\beta)$ is reported \emph{only} when different with respect to $F_{\mbox{\tiny{s}}}^{(1) \mbox{\tiny{num}}}(\beta)$. The sixth, seventh and eighth column display, respectively, the all-loop (full), one-loop (1-l.) and two-loop (2-l.) prediction of the Nambu-Goto string model.}
\label{asymmetric0223102}
\end{table}

\begin{table}[tbp]
\begin{center}
\begin{tabular}{|c|c|c|c|c|c|c|c|}
\hline
$L_{1}$ & $L_{2}$ & $L_{3}$ & $F_{\mbox{\tiny{s}}}^{(1) \mbox{\tiny{num}}}(\beta)$ & $F_{\mbox{\tiny{s}}}^{(2) \mbox{\tiny{num}}}(\beta)$ & $F_{\mbox{\tiny{s}}}^{\mbox{\tiny{th}}}(\beta)$, full & $F_{\mbox{\tiny{s}}}^{\mbox{\tiny{th}}}(\beta)$, 1-l. & $F_{\mbox{\tiny{s}}}^{\mbox{\tiny{th}}}(\beta)$, 2-l. \\
\hline
   24 & 64 &  96 &  7.9988(22)  & 7.9985(22) &  7.9397 &  8.1043 &  7.9802 \\ 
   28 & 64 &  96 &  9.0175(23)  &            &  9.0066 &  9.1016 &  9.0212 \\ 
   32 & 64 &  96 &  9.9965(22)  &            &  9.9962 & 10.0594 & 10.0023 \\ 
   36 & 64 &  96 & 10.9407(23)  &            & 10.9455 & 10.9917 & 10.9482 \\ 
   40 & 64 &  96 & 11.8682(25)  &            & 11.8707 & 11.9070 & 11.8719 \\ 
   44 & 64 &  96 & 12.7728(25)  &            & 12.7803 & 12.8102 & 12.7808 \\ 
   48 & 64 &  96 & 13.6734(27)  &            & 13.6791 & 13.7049 & 13.6793 \\ 
\hline
\end{tabular}
\end{center}
\caption{Same as in table~\ref{asymmetric0223102}, but at $\beta=0.223452$.}
\label{asymmetric0223452}
\end{table}

\begin{table}[tbp]
\begin{center}
\begin{tabular}{|c|c|c|c|c|c|c|}
\hline
$L_{1}$ & $L_{2}$ & $L_{3}$ & $F_{\mbox{\tiny{s}}}^{(2) \mbox{\tiny{num}}}(\beta)$ & $F_{\mbox{\tiny{s}}}^{\mbox{\tiny{th}}}(\beta)$, full & $F_{\mbox{\tiny{s}}}^{\mbox{\tiny{th}}}(\beta)$, 1-l. & $F_{\mbox{\tiny{s}}}^{\mbox{\tiny{th}}}(\beta)$, 2-l. \\
\hline
   24 & 64 &  96 &    9.7374(23)  &  9.7111 &  9.8216 &  9.7302 \\ 
   28 & 64 &  96 &   11.0702(24)  & 11.0653 & 11.1318 & 11.0725 \\ 
   32 & 64 &  96 &   12.3602(24)  & 12.3572 & 12.4024 & 12.3603 \\ 
   36 & 64 &  96 &   13.6150(25)  & 13.6141 & 13.6476 & 13.6155 \\ 
   40 & 64 &  96 &   14.8520(27)  & 14.8492 & 14.8757 & 14.8499 \\ 
   44 & 64 &  96 &   16.0688(27)  & 16.0699 & 16.0918 & 16.0701 \\ 
   48 & 64 &  96 &   17.2816(29)  & 17.2804 & 17.2993 & 17.2804 \\ 
\hline
\end{tabular}
\end{center}
\caption{Same as in table~\ref{asymmetric0223102}, but at $\beta=0.223952$.}
\label{asymmetric0223952}
\end{table}

\begin{table}[tbp]
\begin{center}
\begin{tabular}{|c|c|c|c|c|c|c|}
\hline
$L_{1}$ & $L_{2}$ & $L_{3}$ & $F_{\mbox{\tiny{s}}}^{(2) \mbox{\tiny{num}}}(\beta)$ & $F_{\mbox{\tiny{s}}}^{\mbox{\tiny{th}}}(\beta)$, full & $F_{\mbox{\tiny{s}}}^{\mbox{\tiny{th}}}(\beta)$, 1-l. & $F_{\mbox{\tiny{s}}}^{\mbox{\tiny{th}}}(\beta)$, 2-l. \\
\hline
   24 & 64 &  96 &  18.4131(26) & 18.4121 & 18.4555 & 18.4152 \\ 
   28 & 64 &  96 &  21.2414(27) & 21.2450 & 21.2724 & 21.2463 \\ 
   32 & 64 &  96 &  24.0310(27) & 24.0306 & 24.0497 & 24.0312 \\ 
   36 & 64 &  96 &  26.7859(28) & 26.7873 & 26.8017 & 26.7875 \\ 
   40 & 64 &  96 &  29.5271(30) & 29.5250 & 29.5365 & 29.5251 \\ 
   44 & 64 &  96 &  32.2449(31) & 32.2498 & 32.2594 & 32.2498 \\ 
   48 & 64 &  96 &  34.9623(33) & 34.9653 & 34.9736 & 34.9653 \\ 
\hline
\end{tabular}
\end{center}
\caption{Same as in table~\ref{asymmetric0223102}, but at $\beta=0.226102$.}
\label{asymmetric0226102}
\end{table}

Since we have fixed values of $L_0$, $L_1$ and $L_2$, our data do not allow for a check of scaling corrections. Instead, we assume that they are of similar size as in the case of the square interfaces and therefore are small for the values of $\beta$ that we consider here. 

Similar to the case of square interfaces, the numerical values of the full NG result and the two-loop approximation are very close down to $\sqrt{\sigma} L_1 \approx 1.8$, i.e. they can not be discriminated, given the accuracy of our Monte Carlo results for the Ising model.

In the range $\sqrt{\sigma} L_1  \gtrapprox  1.8$, the Monte Carlo data perfectly agree with the full NG result and the two-loop approximation. Up to our largest values of  $\sqrt{\sigma} L_1$, there is a clear discrimination against the 1-loop approximation.

For $\sqrt{\sigma} L_1  < 1.8$ the Monte Carlo data are more compatible with the 2-loop approximation than with the full NG result.

In the case of the Polyakov-loop correlator, for large distances of the Polyakov-loops, similar observations were made with respect to a finite temperature; i.e. for the direction with periodic boundary conditions ~\cite{Caselle:2005vq, Panero:2005iu, Caselle:2005xy, Panero:2004zq, Caselle:2004jq, Caselle:2003db, Caselle:2003rq, Caselle:2002ah, Caselle:2002vq, Caselle:2002rm}. Even for $\sqrt{\sigma} L_1$ close to $\sqrt{\sigma}/\Tc$, the numerical data follow closely the 2-loop prediction. 

In our opinion, this behaviour at very small distances is a mere coincidence, related to the fact that the 2-loop approximation gives (by chance) the same critical exponent as the 2D Ising universality class and also the value for $\Tc/\sqrt{\sigma}$ is very close to the one of the $\Z_2$ gauge theory.


\section{Universal amplitude ratios}
\label{amplituderatiosect}

In this section we compute universal amplitude ratios  of the interface tension and the correlation length. Generically, these  amplitude ratios have the form:
\begin{equation}
 R = \lim_{t \rightarrow 0}  \sigma(t) \xi(t)^2 \;\; ,
\end{equation}
where $t$ is the reduced temperature. In particular, we shall consider the second moment correlation length in the high temperature and the low temperature phase and the exponential correlation length (i.e. the inverse of the lightest mass) in the low temperature phase.

\subsection{The second moment correlation length in the high temperature phase}

The second moment correlation length is defined by:
\begin{equation}
\xi_{\mbox{\tiny{2nd}}}^2 = \frac{\mu}{2 d \chi} \;\; ,
\end{equation}
where: 
\begin{equation}
\mu= \sum_x x^2  \langle s_0 s_x \rangle
\end{equation}
and $\chi$ is the magnetic susceptibility:
\begin{equation}
\chi= \sum_x  \langle s_0 s_x \rangle \;\;.
\end{equation}
The coefficients of the high temperature series for $\mu$ and $\chi$ up to $O(\beta^{25})$ are reported in ref.~\cite{buco02}. Using these, we computed the coefficients for $\xi_{\mbox{\tiny{2nd}}}^2$. Quite a straightforward method is the so-called matching method (see e.g.~\cite{guttmann}): We start from  a power-law ansatz like:
\begin{eqnarray}
\label{standard1}
&&\xi_{\mbox{\tiny{2nd}}}^2 = f_{\mbox{\tiny{2nd,+}}}^2  (\beta_{\mbox{\tiny{c}}}-\beta)^{-2 \nu} \;\;  \\
&&[1 + a (\beta_{\mbox{\tiny{c}}}-\beta)^{\theta}+b (\beta_{\mbox{\tiny{c}}}-\beta) + 
         c (\beta_{\mbox{\tiny{c}}}-\beta)^{2 \theta} + d (\beta_{\mbox{\tiny{c}}}-\beta)^{\theta_2} + ...
] \;, \nonumber
\end{eqnarray}
where we fix $\beta_{\mbox{\tiny{c}}}=0.22165455$, $\nu=0.6302$, $\theta=0.5174$  and $\gamma=1.2372$, following the Monte Carlo results of ref.~\cite{DB03}. Now the ansatz~(\ref{standard1}) is Taylor-expanded in $\beta$. The remaining free parameters of the ansatz ~(\ref{standard1}) can then be determined by matching the coefficients  with those of the exact result for the HT-series for $\xi_{\mbox{\tiny{2nd}}}^2$. As our final result for the amplitude, we obtain: 
\begin{eqnarray}
\label{xifinal}
 f_{\mbox{\tiny{2nd,+}}}^2 = 0.038369(13) \phantom{xxxxxxxxxxxxxxxxxxxxxxxxxxxxxxxxxxxxx} \\
\times 
[1 + 1024 \; (\beta_{\mbox{\tiny{c}}}-0.22165455) 
 - 17 \; (\nu - 0.6302) -  0.0206 \; (\theta-0.5174) ] \;\;.
\nonumber 
\end{eqnarray}

Combining the results of eqs.~(\ref{sigmafinal},\ref{xifinal}), we get:
\begin{eqnarray}
R_+=0.3869(14)  \phantom{xxxxxxxxxxxxxxxxxxxxxxxxxxxxxxxxxxxxxxx}  \\
+ 1250  (\beta_{\mbox{\tiny{c}}} - 0.22165455)
               + 0.11  (\nu - 0.6302)
               - 0.017   (\theta - 0.5174) \;\;.
\nonumber
\end{eqnarray}
Inserting the errors of the input parameters, we arrive at: 
\begin{equation}
R_+=0.387(2)
\end{equation}
as our final estimate. The most precise theoretical estimate given in the literature $R_+ = 0.377(11)$~\cite{ZinnFisher,ZinnFisher2} is fully consistent with ours. Similar to our analysis, this result is derived from HT-series of the correlation length and Monte Carlo data of the interface tension.

Experimental measurements of this quantity are consistent with but less precise than our estimate; for instance, $R_+=0.41(4)$ was obtained from the study of a cyclohexane-aniline mixture in~\cite{woermann}.

\subsection{The correlation length in the low temperature phase}

Here we use the Monte Carlo results for the second moment correlation length $\xi_{\mbox{\tiny{2nd}}}$ obtained in ref.~\cite{CaHa97} and the exponential correlation length from refs.~\cite{Agostini:1996xy,CaHaPr} to compute the combination $\sigma(t) \xi(t)^2$ at finite values of $t$.  With our present results for the interface tension, the error of the combination is completely dominated by the error of the correlation length.  We fit the combination with the ansatz: 
\begin{equation}
 \sigma \xi^2  =  R_-  \; + \; c \; \sigma^{\omega/2}~~,
\end{equation}
using $\omega=0.821(5)$~\cite{DB03}. In the case of the second moment correlation length we take our final estimate from a fit that includes $\beta \le 0.2275$. The dependence of the result on the value of $\omega$ is rather small and can be neglected compared with the statistical error of $R_-$. In the case of the exponential correlation length our final estimate is obtained from data with $\beta \le 0.23142$ where, again, the error due to the uncertainty of $\omega$ can be ignored. The values of the interface tension, the correlation length and the amplitude combination $R_-$ as well as the final results of our fits are summarized in table~\ref{xis}.

\begin{table}[tbp]
\begin{center}
\begin{tabular}{|c|c|c|c|c|c|}
\hline
\multicolumn{1}{|c}{$\beta$} &
\multicolumn{1}{|c}{$\sigma$} &
\multicolumn{1}{|c}{$\xi_{\mbox{\tiny{exp}}}$} &
\multicolumn{1}{|c}{$ \sigma \xi_{\mbox{\tiny{exp}}}^2$} &
\multicolumn{1}{|c}{$\xi_{\mbox{\tiny{2nd}}}$} &
\multicolumn{1}{|c|}{$\sigma \xi_{\mbox{\tiny{2nd}}}^2$} \\
\hline
0.23910 & 0.055415 &  1.296(3)\phantom{$^*$}     &  0.0931(4)\phantom{0}  & 1.2335(15) & 0.0843(2) \\
0.23142 & 0.027601 &  1.868(3)$^*$ &  0.0964(3)\phantom{0}  & 1.8045(21) & 0.0899(2) \\
0.22750 & 0.014740 &  2.593(3)$^*$ &  0.0991(2)\phantom{0}  & 2.5114(31) & 0.0930(2) \\
0.22600 & 0.010228 &  3.135(9)\phantom{$^*$}     &  0.1005(6)\phantom{0}  & 3.0340(32) & 0.0942(2) \\
0.22400 & 0.004761 &  4.64(3)\phantom{$^*$0}      &  0.1025(13) & 4.509(6)\phantom{00} & 0.0968(3) \\
0.22311 & 0.002626 &     -         &     -       & 6.093(9)\phantom{00}   & 0.0975(3) \\
\hline
$\beta \rightarrow \beta_{\mbox{\tiny{c}}}$ &   &  &  0.1084(11) &            & 0.1024(5) \\
\hline
\end{tabular}
\end{center}
 \caption{
\label{xis}
The values for the interface tension are taken from our global fit, the values for the exponential correlation length are taken from ref.~\cite{Agostini:1996xy} and in the cases marked by $^*$ the results of ref.~\cite{Agostini:1996xy} and ref.~\cite{CaHaPr} are averaged. The numbers for the second moment correlation length are all taken from ref.~\cite{CaHa97}. In the last row we give the extrapolation to the scaling limit. For details see the text.
  }
\end{table}

For comparison, table~\ref{literature} reports previous estimates obtained for the Ising model in refs.~\cite{tension, ZinnFisher, Agostini:1996xy, HaPi97, Klessinger}.

\begin{table}[tbp]
\begin{center}
\begin{tabular}{|c|l|c|l|}
\hline
\multicolumn{1}{|c}{year} &
\multicolumn{1}{|c}{authors(s)} &
\multicolumn{1}{|c}{Ref.} &
\multicolumn{1}{|c|}{$R_{-}$} \\
\hline
1992 & Klessinger and M\"unster & \cite{Klessinger}       & 0.090(3)   \\
1993 & Hasenbusch and Pinn      & \cite{tension}          & 0.090(5)   \\
1996 & Zinn and Fisher          & \cite{ZinnFisher}       & 0.096(2)$^*$   \\
1996 & Agostini \emph{et al.}   & \cite{Agostini:1996xy}  & 0.1056(19) \\
1997 & Hasenbusch and Pinn      & \cite{HaPi97}           & 0.1040(8)$^*$  \\
\hline
\end{tabular}
\end{center} 
\caption{
\label{literature}
  Comparison of a number of estimates for $R_{-}$ taken from the literature. The estimate of Zinn and Fisher is based on data of~\cite{tension}. Agostini \emph{et al.} used the true instead of the second moment correlation length. The result marked with a star refer to the second moment correlation length, while the others refer to the exponential one.}
\end{table}

Also field-theoretical predictions for the universal constant $R$ are given in the literature:
\begin{itemize}
\item From the $\epsilon$-expansion: Br\'ezin and Feng~\cite{BreFeng} computed $R$ from the $\epsilon$-expansion. Their result reads:
\begin{equation}
 \frac{1}{4 \pi R} = \frac{2 \pi }{3} \epsilon \left[1-\epsilon \left(\frac{47}{54} +
 \frac{1}{2} \ln(4\pi) -  \frac{1}{2} \gamma - \frac{5 \pi \sqrt{3}}{18}
 \right)\right] \;+ \; O(\epsilon^{3}) \;\;,
\end{equation}
with $\gamma=0.5772...$.  Their numerical evaluation for $\epsilon=1$ gives results in the range from $R \approx 0.051$ up to $R \approx 0.057$. Note that this result deviates by about a factor of $1/2$ from ours. 
\item 
From perturbation theory in 3D fixed: The study of interfaces using perturbation theory in three dimensions was pioneered by M\"unster~\cite{Munster:1990yg}. The starting point of this calculation is the classical solution (i.e. the configuration with minimal action) of a system with fixed boundary conditions in 3-direction. At one boundary, the field is fixed to the negative minimum, and at the other, to the positive minimum of the potential. Then, fluctuations around this classical solution are studied. In the more recent paper~\cite{HoMu}, this analysis was extended to two loops. Their result\footnote{Note that they use the second moment definition of the correlation length.} is:
\begin{equation}
 R = \frac{2}{u^*_R} \left[ 1 +  \sigma_{1l} \frac{u^*_R}{4\pi} +
                \left(   \sigma_{2l} \frac{u^*_R}{4\pi} \right)^2
+O(u^{*3}_R) \right] \; ,
\end{equation}
with:
\begin{equation}
 \sigma_{1l} = \frac{1}{4} \left(3 + \frac{3}{4} \log 3\right) -
               \frac{37}{32} = - 0.2002602
\end{equation}
and:
\begin{equation}
\sigma_{2l} = -0.0076(8) \;\;.
\end{equation}
Using Pad\'e and Pad\'e-Borel analysis, with $u_R^* = 14.3(1)$, they obtain:
\begin{equation}
R = 0.1065(9) 
\end{equation}
as their final result. Note that the quoted error is dominated by the error of $u_R^*$. At this point, one should also note the principle problems related with the definition of $u_R^*$, as discussed in ref.~\cite{CaHaPr}.
\end{itemize}

Our value $\sigma \xi_{\mbox{\tiny{exp}}}^2 =0.1084(11)$ corresponds to $m_{0++}/\sqrt{\sigma}=3.037(15)$. This value for the $\Z_2$ gauge theory might be compared with $4.718(43)$, $4.329(41)$, $4.236(50)$ and $4.184(55)$ for the $SU(2)$, $SU(3)$, $SU(4)$ and $SU(5)$ gauge theories in 2+1 dimensions, respectively~\cite{JoTe}. Note that there are clear differences between the results for the different gauge groups, indicating that the $0^{++}$ glueball state probes short distances, which
can not be described by an effective string. This is in contrast to higher exited glueball states, where much less dependence on the gauge group is found~\cite{Agostini:1996xy,JoTe}.

In a similar way, we have analysed $\sigma/\Tc^2$, using the results of ref.~\cite{Caselle:1995wn} for $\Tc$.  We arrive at $\sigma/\Tc^2=0.656(2)$ or $\Tc/\sqrt{\sigma}=1.235(2)$, where we have included data with $(1/\Tc) \ge 8$ into the fit. Again the error is completely dominated by the error of $\Tc$.  Our new result can be compared with our previous estimate $\Tc/\sqrt{\sigma}=1.2216(24)$~\cite{Caselle:1995wn}, (where we did not take into account scaling corrections, and had the interface tension only available up to $1/\Tc=12$) and $\Tc/\sqrt{\sigma}=1.17(10)$~\cite{CaGl91}.

Our result is clearly different from the value $\sqrt{3/\pi}=0.977...$ obtained from the effective string picture~\cite{olesen_dec}. 

Finally we might compare with $SU(N)$ gauge theories in $2+1$ dimensions: \\
In the literature one finds $\Tc/\sqrt{\sigma}=1.12(1)$, $0.98(2)$ for $N=2,3$~\cite{teperold} and $\Tc/\sqrt{\sigma}=$ 0.892(3), 0.879(3), 0.877(3) for $N=4,5,6$~\cite{LiTe}, respectively. The results for $N=4,5,6$  are read off from figure 4 of ref.~\cite{LiTe}. Note that they are taken for $N_t=3$ and no continuum extrapolation is performed.

Similar to the case of $m_{0++}/\sqrt{\sigma}$ we observe a clear dependence on the gauge group. Nevertheless, it is quite remarkable that the simple string picture gives the correct value with less than $30 \%$ deviation.

\section{Conclusions}
\label{conclusionsect}

In this paper, we have presented the results of an accurate numerical study of the interface free energy in the three-dimensional Ising model. The motivation for this study was twofold:
\begin{itemize}
\item 
To investigate the dynamics of a fluctuating interface with periodic boundary conditions only, avoiding non-trivial effects of Dirichlet boundary conditions. 

\item To obtain high precision estimates of the interface/string tension in a large range of the inverse temperature, allowing us to compute the scaling limit of several universal amplitude ratios with high precision.

\end{itemize}

To this end, we have determined the interface free energy by numerical integration of the interface energy over the inverse temperature $\beta$. The interface energy is given by the difference of the internal energies in a system with periodic and a system with antiperiodic boundary conditions in one direction. The interface free energy at the starting point of the integration was measured using the boundary-flip algorithm. For our simulations we have used efficient combinations of cluster and multispin coded Metropolis updates. This approach  has allowed us to strongly improve the precision of the numerical results, as compared to analogous studies presented in the literature. In particular, the large range of $\beta$-values that we have studied gives us a good control over corrections to scaling (or, in the language of lattice gauge theory: finite $a$ effects). It turns out that, in the regime not too close to the finite temperature transition, the interface free energy $F_{\mbox{\tiny{s}}}(\sqrt{\sigma} L_1,\sqrt{\sigma} L_2)$ approaches its continuum limit quite fast, characterized by a correction exponent $\omega'\approx 2$. This observation might be explained by the fact that the effective interface model only assumes the restoration of the symmetries of the continuous space-time, which indeed comes with an exponent $\omega'\approx 2$. 

The level of precision in the Monte Carlo data, as well as the accurate control of the  systematic errors, enables us to clearly resolve the fine string-dynamics effects that we were seeking after: In a setting in which the possible distortions due to boundary effects are completely absent, and all systematic effects are under control, the Nambu-Goto model yields an accurate description of the data up to the second loop order (only). 

Comparing the 2-loop approximation with the full NG prediction one has to notice that down to rather small scales, such as $\sqrt{\sigma} L \approx 1.8$, the two string results can not be discriminated at the level of the accuracy of our Monte Carlo data. Nevertheless, one might interpret our Monte Carlo data as a confirmation of the full NG prediction in the sense that e.g. $1/(\sigma A)^2$ corrections have indeed a small amplitude.

Going close to the finite temperature transition, which occurs at $\sqrt{\sigma} N_t =0.810(2)$ as discussed in section~\ref{amplituderatiosect}, it does not come as too big a surprise that the data for the Ising model are not well fitted by the full NG prediction: The full NG predicts mean-field behaviour of the transition, while 
following the Svetitsky-Yaffe conjecture~\cite{sy82} (numerically confirmed e.g. in~\cite{Caselle:1995wn}) it should be the behaviour of the 2D Ising universality class.

This behaviour is quite similar to that of interfaces with a cylinder-like geometry, with periodic boundary conditions in the short direction (see e.g. ref.~\cite{Caselle:2002rm}), thus confirming the assumption that a common effective string description underlies the confining flux tube dynamics in different physical settings. 

These observations indicate that the all-order prediction of the Nambu-Goto effective string action (at least as it is treated in the approximation in which the r\^ole of the Liouville field is neglected) does not show a quantitative agreement with the data for an interface in the three-dimensional Ising model that goes beyond the two-loop approximation. This fact is --- as discussed above --- consistent with the Polchinski-Strominger model.

The apparent failure of the two-loop prediction with respect to Dirichlet boundary conditions (see e.g. ref. \cite{Caselle:2004jq}), as they arise via duality from the Polyakov-loop correlator requires further investigation. At least the present work confirms that the problem is intrinsically related with the boundary conditions. A possible explanation is that the Dirichlet boundary conditions probe short distance properties of the theory, which are not captured by the effective string model.

Using our new data for the interface tension along with the analysis of the high temperature series expansion of the second moment correlation length, we have computed the universal amplitude ratio $R_+ = 0.387(2)$. This estimate is more precise than any other theoretical estimate given in the literature. This estimate can be compared with experimental results e.g. for binary mixtures. In ref.~\cite{woermann} $R_+=0.41(4)$ was found, which is consistent but less precise than our result. 

Finally, we have also updated the results for the mass $m_{0++}/\sqrt{\sigma}$ of the $0^{++}$ glueball in units of the square root of the string tension and the critical temperature of the deconfinement transition $\Tc/\sqrt{\sigma}$. 

The comparison with other gauge theories in $2+1$ dimensions shows a variation of these quantities by roughly $50\%$, indicating that these quantities can not be described only by an effective string picture, but that also microscopic features of the gauge theory have to be taken into account.

\vskip1.0cm \noindent {\bf
Acknowledgements.}

\noindent M.P. acknowledges support from the Alexander von Humboldt Foundation.

\appendix{}
\vskip 0.5cm

\section{Integration schemes}
\label{integrationschemesect}

\renewcommand{\theequation}{A.\arabic{equation}}
\setcounter{equation}{0}

One of the simplest methods given in textbooks is the trapezoid rule: \\
\begin{equation}
\label{trapez}
\int_{x_0}^{x_N} f(x) \mbox{d} x =  h \left[
\frac{1}{2} f_0 +  f_1 +  f_2 + ... + f_{N-1} + \frac{1}{2} f_N \right]
+O(N^{-2})\;,
\end{equation}
where $h=(x_N-x_0)/N$, $f_i = f(x_i)$ and $x_i=x_0+i h$. There are rules with faster convergence as $N \rightarrow \infty$, like the well-known Simpson rule:
\begin{eqnarray}
\int_{x_0}^{x_N} f(x) \mbox{d} x =  h \left[
\frac{1}{3} f_0 + \frac{4}{3} f_1 + \frac{2}{3} f_2 + \frac{4}{3} f_3 + 
\frac{2}{3} f_4 + ...
\phantom{xxxxxxxxx} \right . \nonumber \\
\phantom{xxxxxxxxxxxxx}
\left . + \frac{4}{3} f_{N-3} +
\frac{2}{3} f_{N-2} + \frac{4}{3} f_{N-1} + \frac{1}{3} f_N \right]
+O(N^{-4}) \;\;.
\end{eqnarray}
However, the disadvantage of the Simpson rule is that $f_{i}$ is not constant in the middle of the interval: this implies a loss of precision in the final result. Furthermore, $N$ has to be even.

A better-suited rule that avoids these problems is given by (see e.g. eq.~(4.1.14) in ref.~\cite{NUMRES}):
\begin{eqnarray}
\label{n4x1}
\int_{x_0}^{x_N} f(x) \mbox{d} x =  h \left[
\frac{3}{8} f_0 + \frac{7}{6} f_1 + \frac{23}{24} f_2 + f_3 + f_4 + ... 
\phantom{xxxxxxxxxxxx} \right . \nonumber \\
\phantom{xxxxxxxxxxxx}
\left . + f_{N-3} +
\frac{23}{24} f_{N-2} + \frac{7}{6} f_{N-1} + \frac{3}{8} f_N \right]
+O(N^{-4}).
\end{eqnarray}

A similar and maybe slightly better rule (see e.g. eq.~(35) in ref.~\cite{wolfram}) is given by:
\begin{eqnarray}
\label{n4x2}
\int_{x_0}^{x_N} f(x) \mbox{d} x =  h \left[
\frac{17}{48} f_0 + \frac{59}{48} f_1 + \frac{43}{48} f_2 + \frac{49}{48} f_3 + f_4 + ...
\phantom{xxxxxxxxx} \right . \nonumber \\
\phantom{xxxxx}
\left .+f_{N-4} + \frac{49}{48} f_{N-3} +
\frac{43}{48} f_{N-2} + \frac{59}{48} f_{N-1} + \frac{17}{48} f_N \right]
+O(N^{-4}).
\end{eqnarray}

\end{document}